\documentclass[onecolumn]{aa} 

\usepackage{graphicx, amsmath}

\usepackage{natbib}

\usepackage{txfonts}
\usepackage{upgreek}
\usepackage{lineno}
\usepackage[encapsulated]{CJK}

\usepackage{amsmath}
\usepackage{float}
\usepackage[T1]{fontenc}
\usepackage[utf8]{inputenc}
\usepackage{amssymb}
\usepackage{textcomp}
\usepackage{gensymb}
\usepackage[procnames]{listings}
\usepackage{tabto}
\usepackage{tabularx}
\usepackage{paralist} 
\usepackage{mathtools}

\usepackage[normalem]{ulem}
\usepackage{xcolor}
\usepackage{xspace}
\usepackage{comment}
\usepackage{ucs}

\usepackage[nolist,nohyperlinks]{acronym}

\usepackage{siunitx}
\usepackage{changepage}
\usepackage[colorlinks=true]{hyperref}
\hypersetup{
     colorlinks   = true,
     citecolor    = blue
}

\newcommand{\C}{3C\,84}
\newcommand{\Tbturn}{(3.6\pm1.5)\times10^{11}\,\textrm{K}} 
\newcommand{\Tbturnnoe}{3.6\times10^{11}\,\textrm{K}} 
\newcommand{\Bssa}{(2.9\pm1.6)\,\textrm{G}}
\newcommand{\Beq}{(5.2\pm0.6)\,\textrm{G}} 
\newcommand{\numf}{(113\pm4)\,\textrm{GHz}} 
\newcommand{\mtot}{(17.0\pm3.9)\%} 
\newcommand{\frm}{(6.06\pm0.01)\times10^6\,\textrm{rad/m}^2}

\newcommand{\phim}{(144\pm18)\,\upmu\textrm{as}}
\newcommand{\magpar}{41-93} 
\newcommand{\deq}{1.5\pm0.4} 
\newcommand\sbullet[1][.5]{\mathbin{\vcenter{\hbox{\scalebox{#1}{$\bullet$}}}}}
\newcommand{\Sm}{9.0\pm0.5\,\textrm{Jy}} 
\newcommand{\Szero}{5.7\pm0.3\,\textrm{Jy}}
\newcommand{\Score}{6.4\,\textrm{Jy}}
\newcommand{\Baver}{4.7\pm0.6\,\textrm{mG}}

\newcommand{\cntext}[1]{\begin{CJK}{UTF8}{gbsn}#1\end{CJK}}

\begin{document}

    \title{Ordered magnetic fields around the \C\ central black hole}

   \author{G.~F. Paraschos\inst{1}, J.-Y. Kim\inst{2,1}, M. Wielgus\inst{1,3}, J. Röder\inst{1}, T.~P. Krichbaum\inst{1}, E. Ros\inst{1}, I. Agudo\inst{4}, I. Myserlis\inst{5,1}, M. Moscibrodzka\inst{6}, E. Traianou\inst{4, 1}, J.~A. Zensus\inst{1}
% ...
   }

   \authorrunning{G.F. Paraschos et al.}
   
\institute{
$^1$ Max-Planck-Institut für Radioastronomie, Auf dem Hügel 69, D-53121 Bonn, Germany\\
$^{}$\ \email{gfparaschos@mpifr-bonn.mpg.de}\\
$^2$ Department of Astronomy and Atmospheric Sciences, Kyungpook National University, Daegu 702-701, Republic of Korea\\
$^3$ Institute of Physics, Silesian University in Opava, Bezru\v{c}ovo n\'{a}m. 13, CZ-746 01 Opava, Czech Republic\\
$^4$ Instituto de Astrofísica de Andalucía-CSIC, Glorieta de la Astronomía s/n, E-18008 Granada, Spain\\
$^5$ Institut de Radioastronomie Millimétrique (IRAM), Avenida Divina Pastora 7, Local 20, E-18012, Granada, Spain\\
$^6$ Department of Astrophysics, Institute for Mathematics, Astrophysics and Particle Physics (IMAPP), Radboud University, P.O. Box 9010, 6500 GL Nijmegen, The Netherlands\\
$^7$ Black Hole Initiative at Harvard University, 20 Garden Street, Cambridge, MA 02138, USA\\
$^8$ Center for Astrophysics $|$ Harvard \& Smithsonian, 60 Garden Street, Cambridge, MA 02138, USA\\
$^9$ Steward Observatory and Department of Astronomy, University of Arizona, 933 N. Cherry Ave., Tucson, AZ 85721, USA\\
$^{10}$ Data Science Institute, University of Arizona, 1230 N. Cherry Ave., Tucson, AZ 85721, USA\\
$^{11}$ Program in Applied Mathematics, University of Arizona, 617 N. Santa Rita, Tucson, AZ 85721, USA\\
$^{12}$ NASA Hubble Fellowship Program, Einstein Fellow\\
$^{13}$ Massachusetts Institute of Technology Haystack Observatory, 99 Millstone Road, Westford, MA 01886, USA\\
$^{14}$ National Astronomical Observatory of Japan, 2-21-1 Osawa, Mitaka, Tokyo 181-8588, Japan\\
$^{15}$ Department of Physics, Faculty of Science, Universiti Malaya, 50603 Kuala Lumpur, Malaysia\\
$^{16}$ Department of Physics \& Astronomy, The University of Texas at San Antonio, One UTSA Circle, San Antonio, TX 78249, USA\\
$^{17}$ Institute of Astronomy and Astrophysics, Academia Sinica, 11F of Astronomy-Mathematics Building, AS/NTU No. 1, Sec. 4, Roosevelt Rd., Taipei 10617, Taiwan, R.O.C.\\
$^{18}$ Departament d'Astronomia i Astrofísica, Universitat de València, C. Dr. Moliner 50, E-46100 Burjassot, València, Spain\\
$^{19}$ Observatori Astronòmic, Universitat de València, C. Catedrático José Beltrán 2, E-46980 Paterna, València, Spain\\
$^{20}$ Department of Space, Earth and Environment, Chalmers University of Technology, Onsala Space Observatory, SE-43992 Onsala, Sweden\\
$^{21}$ Yale Center for Astronomy \& Astrophysics, Yale University, 52 Hillhouse Avenue, New Haven, CT 06511, USA\\
$^{22}$ Department of Physics, University of Illinois, 1110 West Green Street, Urbana, IL 61801, USA\\
$^{23}$ Fermi National Accelerator Laboratory, MS209, P.O. Box 500, Batavia, IL 60510, USA\\
$^{24}$ Department of Astronomy and Astrophysics, University of Chicago, 5640 South Ellis Avenue, Chicago, IL 60637, USA\\
$^{25}$ East Asian Observatory, 660 N. A'ohoku Place, Hilo, HI 96720, USA\\
$^{26}$ James Clerk Maxwell Telescope (JCMT), 660 N. A'ohoku Place, Hilo, HI 96720, USA\\
$^{27}$ California Institute of Technology, 1200 East California Boulevard, Pasadena, CA 91125, USA\\
$^{28}$ Institute of Astronomy and Astrophysics, Academia Sinica, 645 N. A'ohoku Place, Hilo, HI 96720, USA\\
$^{29}$ Department of Physics and Astronomy, University of Hawaii at Manoa, 2505 Correa Road, Honolulu, HI 96822, USA\\
$^{30}$ Department of Physics, McGill University, 3600 rue University, Montréal, QC H3A 2T8, Canada\\
$^{31}$ Trottier Space Institute at McGill, 3550 rue University, Montréal, QC H3A 2A7, Canada\\
$^{32}$ Institut de Radioastronomie Millimétrique (IRAM), 300 rue de la Piscine, F-38406 Saint Martin d'Hères, France\\
$^{33}$ Perimeter Institute for Theoretical Physics, 31 Caroline Street North, Waterloo, ON N2L 2Y5, Canada\\
$^{34}$ Department of Physics and Astronomy, University of Waterloo, 200 University Avenue West, Waterloo, ON N2L 3G1, Canada\\
$^{35}$ Waterloo Centre for Astrophysics, University of Waterloo, Waterloo, ON N2L 3G1, Canada\\
$^{36}$ Department of Astronomy, University of Massachusetts, Amherst, MA 01003, USA\\
$^{37}$ Korea Astronomy and Space Science Institute, Daedeok-daero 776, Yuseong-gu, Daejeon 34055, Republic of Korea\\
$^{38}$ University of Science and Technology, Gajeong-ro 217, Yuseong-gu, Daejeon 34113, Republic of Korea\\
$^{39}$ Kavli Institute for Cosmological Physics, University of Chicago, 5640 South Ellis Avenue, Chicago, IL 60637, USA\\
$^{40}$ Department of Physics, University of Chicago, 5720 South Ellis Avenue, Chicago, IL 60637, USA\\
$^{41}$ Enrico Fermi Institute, University of Chicago, 5640 South Ellis Avenue, Chicago, IL 60637, USA\\
$^{42}$ Princeton Gravity Initiative, Jadwin Hall, Princeton University, Princeton, NJ 08544, USA\\
$^{43}$ Cornell Center for Astrophysics and Planetary Science, Cornell University, Ithaca, NY 14853, USA\\
$^{44}$ Shanghai Astronomical Observatory, Chinese Academy of Sciences, 80 Nandan Road, Shanghai 200030, People's Republic of China\\
$^{45}$ Key Laboratory of Radio Astronomy, Chinese Academy of Sciences, Nanjing 210008, People's Republic of China\\
$^{46}$ Physics Department, Fairfield University, 1073 North Benson Road, Fairfield, CT 06824, USA\\
$^{47}$ Department of Astronomy, University of Illinois at Urbana-Champaign, 1002 West Green Street, Urbana, IL 61801, USA\\
$^{48}$ Instituto de Astronomía, Universidad Nacional Autónoma de México (UNAM), Apdo Postal 70-264, Ciudad de México, México\\
$^{49}$ Institut für Theoretische Physik, Goethe-Universität Frankfurt, Max-von-Laue-Straße 1, D-60438 Frankfurt am Main, Germany\\
$^{50}$ Research Center for Intelligent Computing Platforms, Zhejiang Laboratory, Hangzhou 311100, China\\
$^{51}$ Tsung-Dao Lee Institute, Shanghai Jiao Tong University, Shengrong Road 520, Shanghai, 201210, People’s Republic of China\\
$^{52}$ Department of Astronomy and Columbia Astrophysics Laboratory, Columbia University, 500 W. 120th Street, New York, NY 10027, USA\\
$^{53}$ Center for Computational Astrophysics, Flatiron Institute, 162 Fifth Avenue, New York, NY 10010, USA\\
$^{54}$ Dipartimento di Fisica ``E. Pancini'', Università di Napoli ``Federico II'', Compl. Univ. di Monte S. Angelo, Edificio G, Via Cinthia, I-80126, Napoli, Italy\\
$^{55}$ INFN Sez. di Napoli, Compl. Univ. di Monte S. Angelo, Edificio G, Via Cinthia, I-80126, Napoli, Italy\\
$^{56}$ Wits Centre for Astrophysics, University of the Witwatersrand, 1 Jan Smuts Avenue, Braamfontein, Johannesburg 2050, South Africa\\
$^{57}$ Department of Physics, University of Pretoria, Hatfield, Pretoria 0028, South Africa\\
$^{58}$ Centre for Radio Astronomy Techniques and Technologies, Department of Physics and Electronics, Rhodes University, Makhanda 6140, South Africa\\
$^{59}$ ASTRON, Oude Hoogeveensedijk 4, 7991 PD Dwingeloo, The Netherlands\\
$^{60}$ LESIA, Observatoire de Paris, Université PSL, CNRS, Sorbonne Université, Université de Paris, 5 place Jules Janssen, F-92195 Meudon, France\\
$^{61}$ JILA and Department of Astrophysical and Planetary Sciences, University of Colorado, Boulder, CO 80309, USA\\
$^{62}$ National Astronomical Observatories, Chinese Academy of Sciences, 20A Datun Road, Chaoyang District, Beijing 100101, PR China\\
$^{63}$ Las Cumbres Observatory, 6740 Cortona Drive, Suite 102, Goleta, CA 93117-5575, USA\\
$^{64}$ Department of Physics, University of California, Santa Barbara, CA 93106-9530, USA\\
$^{65}$ National Radio Astronomy Observatory, 520 Edgemont Road, Charlottesville, VA 22903, USA\\
$^{66}$ Department of Electrical Engineering and Computer Science, Massachusetts Institute of Technology, 32-D476, 77 Massachusetts Ave., Cambridge, MA 02142, USA\\
$^{67}$ Google Research, 355 Main St., Cambridge, MA 02142, USA\\
$^{68}$ Institut für Theoretische Physik und Astrophysik, Universität Würzburg, Emil-Fischer-Str. 31, D-97074 Würzburg, Germany\\
$^{69}$ Department of History of Science, Harvard University, Cambridge, MA 02138, USA\\
$^{70}$ Department of Physics, Harvard University, Cambridge, MA 02138, USA\\
$^{71}$ NCSA, University of Illinois, 1205 W. Clark St., Urbana, IL 61801, USA\\
$^{72}$ Instituto de Astronomia, Geofísica e Ciências Atmosféricas, Universidade de São Paulo, R. do Matão, 1226, São Paulo, SP 05508-090, Brazil\\
$^{73}$ Dipartimento di Fisica, Università degli Studi di Cagliari, SP Monserrato-Sestu km 0.7, I-09042 Monserrato (CA), Italy\\
$^{74}$ INAF - Osservatorio Astronomico di Cagliari, via della Scienza 5, I-09047 Selargius (CA), Italy\\
$^{75}$ INFN, sezione di Cagliari, I-09042 Monserrato (CA), Italy\\
$^{76}$ CP3-Origins, University of Southern Denmark, Campusvej 55, DK-5230 Odense M, Denmark\\
$^{77}$ Instituto Nacional de Astrofísica, Óptica y Electrónica. Apartado Postal 51 y 216, 72000. Puebla Pue., México\\
$^{78}$ Consejo Nacional de Ciencia y Tecnologìa, Av. Insurgentes Sur 1582, 03940, Ciudad de México, México\\
$^{79}$ Key Laboratory for Research in Galaxies and Cosmology, Chinese Academy of Sciences, Shanghai 200030, People's Republic of China\\
$^{80}$ Mizusawa VLBI Observatory, National Astronomical Observatory of Japan, 2-12 Hoshigaoka, Mizusawa, Oshu, Iwate 023-0861, Japan\\
$^{81}$ Department of Astronomical Science, The Graduate University for Advanced Studies (SOKENDAI), 2-21-1 Osawa, Mitaka, Tokyo 181-8588, Japan\\
$^{82}$ Trottier Space Institute at McGill, 3550 rue University, Montréal,  QC H3A 2A7, Canada\\
$^{83}$ NOVA Sub-mm Instrumentation Group, Kapteyn Astronomical Institute, University of Groningen, Landleven 12, 9747 AD Groningen, The Netherlands\\
$^{84}$ Department of Astronomy, School of Physics, Peking University, Beijing 100871, People's Republic of China\\
$^{85}$ Kavli Institute for Astronomy and Astrophysics, Peking University, Beijing 100871, People's Republic of China\\
$^{86}$ Department of Astronomy, Graduate School of Science, The University of Tokyo, 7-3-1 Hongo, Bunkyo-ku, Tokyo 113-0033, Japan\\
$^{87}$ The Institute of Statistical Mathematics, 10-3 Midori-cho, Tachikawa, Tokyo, 190-8562, Japan\\
$^{88}$ Department of Statistical Science, The Graduate University for Advanced Studies (SOKENDAI), 10-3 Midori-cho, Tachikawa, Tokyo 190-8562, Japan\\
$^{89}$ Kavli Institute for the Physics and Mathematics of the Universe, The University of Tokyo, 5-1-5 Kashiwanoha, Kashiwa, 277-8583, Japan\\
$^{90}$ Leiden Observatory, Leiden University, Postbus 2300, 9513 RA Leiden, The Netherlands\\
$^{91}$ ASTRAVEO LLC, PO Box 1668, Gloucester, MA 01931\\
$^{92}$ Applied Materials Inc., 35 Dory Road, Gloucester, MA 01930\\
$^{93}$ Institute for Astrophysical Research, Boston University, 725 Commonwealth Ave., Boston, MA 02215, USA\\
$^{94}$ Institute for Cosmic Ray Research, The University of Tokyo, 5-1-5 Kashiwanoha, Kashiwa, Chiba 277-8582, Japan\\
$^{95}$ Joint Institute for VLBI ERIC (JIVE), Oude Hoogeveensedijk 4, 7991 PD Dwingeloo, The Netherlands\\
$^{96}$ Kogakuin University of Technology \& Engineering, Academic Support Center, 2665-1 Nakano, Hachioji, Tokyo 192-0015, Japan\\
$^{97}$ Graduate School of Science and Technology, Niigata University, 8050 Ikarashi 2-no-cho, Nishi-ku, Niigata 950-2181, Japan\\
$^{98}$ Physics Department, National Sun Yat-Sen University, No. 70, Lien-Hai Road, Kaosiung City 80424, Taiwan, R.O.C.\\
$^{99}$ National Optical Astronomy Observatory, 950 N. Cherry Ave., Tucson, AZ 85719, USA\\
$^{100}$ Department of Physics, The Chinese University of Hong Kong, Shatin, N. T., Hong Kong\\
$^{101}$ School of Astronomy and Space Science, Nanjing University, Nanjing 210023, People's Republic of China\\
$^{102}$ Key Laboratory of Modern Astronomy and Astrophysics, Nanjing University, Nanjing 210023, People's Republic of China\\
$^{103}$ INAF-Istituto di Radioastronomia, Via P. Gobetti 101, I-40129 Bologna, Italy\\
$^{104}$ INAF-Istituto di Radioastronomia \& Italian ALMA Regional Centre, Via P. Gobetti 101, I-40129 Bologna, Italy\\
$^{105}$ Department of Physics, National Taiwan University, No. 1, Sec. 4, Roosevelt Rd., Taipei 10617, Taiwan, R.O.C\\
$^{106}$ Instituto de Radioastronomía y Astrofísica, Universidad Nacional Autónoma de México, Morelia 58089, México\\
$^{107}$ Yunnan Observatories, Chinese Academy of Sciences, 650011 Kunming, Yunnan Province, People's Republic of China\\
$^{108}$ Center for Astronomical Mega-Science, Chinese Academy of Sciences, 20A Datun Road, Chaoyang District, Beijing, 100012, People's Republic of China\\
$^{109}$ Key Laboratory for the Structure and Evolution of Celestial Objects, Chinese Academy of Sciences, 650011 Kunming, People's Republic of China\\
$^{110}$ Anton Pannekoek Institute for Astronomy, University of Amsterdam, Science Park 904, 1098 XH, Amsterdam, The Netherlands\\
$^{111}$ Gravitation and Astroparticle Physics Amsterdam (GRAPPA) Institute, University of Amsterdam, Science Park 904, 1098 XH Amsterdam, The Netherlands\\
$^{112}$ Department of Astrophysical Sciences, Peyton Hall, Princeton University, Princeton, NJ 08544, USA\\
$^{113}$ Science Support Office, Directorate of Science, European Space Research and Technology Centre (ESA/ESTEC), Keplerlaan 1, 2201 AZ Noordwijk, The Netherlands\\
$^{114}$ School of Physics and Astronomy, Shanghai Jiao Tong University, 800 Dongchuan Road, Shanghai, 200240, People’s Republic of China\\
$^{115}$ Astronomy Department, Universidad de Concepción, Casilla 160-C, Concepción, Chile\\
$^{116}$ National Institute of Technology, Hachinohe College, 16-1 Uwanotai, Tamonoki, Hachinohe City, Aomori 039-1192, Japan\\
$^{117}$ Research Center for Astronomy, Academy of Athens, Soranou Efessiou 4, 115 27 Athens, Greece\\
$^{118}$ Department of Physics, Villanova University, 800 Lancaster Avenue, Villanova, PA 19085, USA\\
$^{119}$ Physics Department, Washington University, CB 1105, St. Louis, MO 63130, USA\\
$^{120}$ School of Physics, Georgia Institute of Technology, 837 State St NW, Atlanta, GA 30332, USA\\
$^{121}$ Department of Astronomy and Space Science, Kyung Hee University, 1732, Deogyeong-daero, Giheung-gu, Yongin-si, Gyeonggi-do 17104, Republic of Korea\\
$^{122}$ Canadian Institute for Theoretical Astrophysics, University of Toronto, 60 St. George Street, Toronto, ON M5S 3H8, Canada\\
$^{123}$ Dunlap Institute for Astronomy and Astrophysics, University of Toronto, 50 St. George Street, Toronto, ON M5S 3H4, Canada\\
$^{124}$ Canadian Institute for Advanced Research, 180 Dundas St West, Toronto, ON M5G 1Z8, Canada\\
$^{125}$ Radio Astronomy Laboratory, University of California, Berkeley, CA 94720, USA\\
$^{126}$ Institute of Astrophysics, Foundation for Research and Technology - Hellas, Voutes, 7110 Heraklion, Greece\\
$^{127}$ Department of Physics, National Taiwan Normal University, No. 88, Sec. 4, Tingzhou Rd., Taipei 116, Taiwan, R.O.C.\\
$^{128}$ Center of Astronomy and Gravitation, National Taiwan Normal University, No. 88, Sec. 4, Tingzhou Road, Taipei 116, Taiwan, R.O.C.\\
$^{129}$ Finnish Centre for Astronomy with ESO, FI-20014 University of Turku, Finland\\
$^{130}$ Aalto University Metsähovi Radio Observatory, Metsähovintie 114, FI-02540 Kylmälä, Finland\\
$^{131}$ Gemini Observatory/NSF NOIRLab, 670 N. A’ohōkū Place, Hilo, HI 96720, USA\\
$^{132}$ Frankfurt Institute for Advanced Studies, Ruth-Moufang-Strasse 1, D-60438 Frankfurt, Germany\\
$^{133}$ School of Mathematics, Trinity College, Dublin 2, Ireland\\
$^{134}$ Department of Physics, University of Toronto, 60 St. George Street, Toronto, ON M5S 1A7, Canada\\
$^{135}$ Department of Physics, Tokyo Institute of Technology, 2-12-1 Ookayama, Meguro-ku, Tokyo 152-8551, Japan\\
$^{136}$ Hiroshima Astrophysical Science Center, Hiroshima University, 1-3-1 Kagamiyama, Higashi-Hiroshima, Hiroshima 739-8526, Japan\\
$^{137}$ Aalto University Department of Electronics and Nanoengineering, PL 15500, FI-00076 Aalto, Finland\\
$^{138}$ Department of Astronomy, Yonsei University, Yonsei-ro 50, Seodaemun-gu, 03722 Seoul, Republic of Korea\\
$^{139}$ National Biomedical Imaging Center, Peking University, Beijing 100871, People’s Republic of China\\
$^{140}$ College of Future Technology, Peking University, Beijing 100871, People’s Republic of China\\
$^{141}$ Department of Physics and Astronomy, University of Lethbridge, Lethbridge, Alberta T1K 3M4, Canada\\
$^{142}$ Netherlands Organisation for Scientific Research (NWO), Postbus 93138, 2509 AC Den Haag, The Netherlands\\
$^{143}$ Department of Physics and Astronomy, Seoul National University, Gwanak-gu, Seoul 08826, Republic of Korea\\
$^{144}$ University of New Mexico, Department of Physics and Astronomy, Albuquerque, NM 87131, USA\\
$^{145}$ Jeremiah Horrocks Institute, University of Central Lancashire, Preston PR1 2HE, UK\\
$^{146}$ Physics Department, Brandeis University, 415 South Street, Waltham, MA 02453, USA\\
$^{147}$ Tuorla Observatory, Department of Physics and Astronomy, University of Turku, Finland\\
$^{148}$ Radboud Excellence Fellow of Radboud University, Nijmegen, The Netherlands\\
$^{149}$ School of Natural Sciences, Institute for Advanced Study, 1 Einstein Drive, Princeton, NJ 08540, USA\\
$^{150}$ School of Physics, Huazhong University of Science and Technology, Wuhan, Hubei, 430074, People's Republic of China\\
$^{151}$ Mullard Space Science Laboratory, University College London, Holmbury St. Mary, Dorking, Surrey, RH5 6NT, UK\\
$^{152}$ School of Astronomy and Space Sciences, University of Chinese Academy of Sciences, No. 19A Yuquan Road, Beijing 100049, People's Republic of China\\
$^{153}$ Astronomy Department, University of Science and Technology of China, Hefei 230026, People's Republic of China\\
$^{154}$ Department of Physics and Astronomy, Michigan State University, 567 Wilson Rd, East Lansing, MI 48824, USA\\
}

   \date{Received -; accepted -}

 \abstract
   {
   \C\ is a nearby radio source with a complex total intensity structure, showing linear polarisation and spectral patterns. 
   A~detailed investigation of the central engine region necessitates the use of very-long-baseline interferometry (VLBI) above the hitherto available maximum frequency of 86\,GHz.
   }
   {
   Using ultrahigh resolution VLBI observations at the currently highest available frequency of 228\,GHz, we aim to perform a direct detection of compact structures and understand the physical conditions in the compact region of \C.
   }
   {
   We used Event Horizon Telescope (EHT) 228\,GHz observations and, given the limited $(u, v)$-coverage, applied geometric model fitting to the data. 
   Furthermore, we employed quasi-simultaneously observed, ancillary multi-frequency VLBI data for the source in order to carry out a comprehensive analysis of the core structure.
   }
   {
    We report the detection of a highly ordered, strong magnetic field around the central, supermassive black hole of \C.
    The brightness temperature analysis suggests that the system is in equipartition. 
    We also determined a turnover frequency of $\nu_\textrm{m}=\numf$, a corresponding synchrotron self-absorbed magnetic field of $B_\textrm{SSA} = \Bssa$, and an equipartition magnetic field of $B_\textrm{eq} = \Beq$.
    Three components are resolved with the highest fractional polarisation detected for this object ($m_\textrm{net} = \mtot$).
    The positions of the components are compatible with those seen in low-frequency VLBI observations since 2017-2018.
    We report a steeply negative slope of the spectrum at 228\,GHz.
    We used these findings to test existing models of jet formation, propagation, and Faraday rotation in \C.
   }
   {
    The findings of our investigation into different flow geometries and black hole spins support an advection-dominated accretion flow in a magnetically arrested state around a rapidly rotating supermassive black hole as a model of the jet-launching system in the core of \C. 
    However, systematic uncertainties due to the limited $(u, v)$-coverage, however, cannot be ignored.
    Our upcoming work using new EHT data, which offer full imaging capabilities, will shed more light on the compact region of \C.
   }


   \maketitle
   \twocolumn

\section{Introduction}

The formation of relativistic astrophysical jets is a manifestation of the activity of accreting supermassive black holes residing in the nuclei of galaxies.
Such jets can have an immense impact on their surroundings, either by stunting or enhancing the evolution of their host galaxy.
Despite substantial efforts dedicated to understanding the physics governing jets, a number of open questions remain, including questions relating to the launching mechanism of these jets.
The radio source \C\ (NGC\,1275; $D_\textrm{L} = 78.9\pm2.4\,\textrm{Mpc}$, $z = 0.0176$, \citealt{Strauss92}, corresponding to a conversion factor $\uppsi=0.36$\,pc/mas; see also Sect.~\ref{ssec:Data}) is a nearby active galactic nucleus (AGN) and one of a handful of objects for which the jet formation zone can be resolved and probed with very-long-baseline interferometry (VLBI).
Thus, \C\ is an ideal test bed for distinguishing between jet-launching models based on the resulting predictions for observables such as magnetic field strength.
Using the unique polarimetric 1.3\,mm VLBI observations of \C, conducted with the Event Horizon Telescope (EHT; see \citealt{EHT19a,EHT22a}), we are now able to distinguish between such models.

According to the current understanding, the linear polarisation is present in both the downstream jet \citep{Nagai17} and the compact region \citep{Kim19} of \C, although its amplitude is low. 
A quantitative characterisation of the location of the 1.3\,mm polarisation within the jet flow is crucial in order to distinguish between the different jet-launching models.
To illustrate this, an interesting comparison can be made between the jet collimation near the jet base in M\,87 (exhibiting a narrower opening angle, as seen, e.g. in \citealt{Kim18}) and \C\ (featuring instead a wide structure as seen by {\it RadioAstron} and reported in \citealt{Giovannini18}).
Given this elongated structure, a disc-launched jet \citep{Blandford82} threaded by toroidal magnetic field lines is a possible explanation. 
The alternative scenario is the more commonly invoked black hole launched jet \citep{Blandford77} associated with poloidal magnetic field lines.
Polarimetry at 1.3\,mm is less affected by opacity effects and can therefore be used to test the necessary conditions for different jet-launching scenarios, as presented in this work. 
We therefore employed high-resolution millimetre VLBI to investigate how the substantial increase in polarisation with frequency in \C\ can be explained by the prevalent magnetic field.

\section{Data, analysis, and results}\label{sec:Results}

\subsection{Data description and analysis}\label{ssec:Data}

In this work, we examined the first total intensity and polarimetric VLBI observations of \C\ at 228\,GHz taken with the Event Horizon Telescope (EHT) and compared them with quasi-simultaneous VLBI observations at lower frequencies.
\C\ was observed during the EHT 2017 campaign \citep{EHT19a,EHT22a} at 228\,GHz on April 7 between 18:30 and 19:40 UTC, with six scans each around 5 mins in length. 
Five telescopes at three geographical sites participated in the observation: Atacama Large Millimeter/submillimeter Array \citep[ALMA, observing as a phased array; see][]{Goddi19} and the Atacama Pathfinder Experiment (APEX) telescope in Chile; the Submillimeter Telescope (SMT) in Arizona; and the James Clerk Maxwell Telescope (JCMT) and the Submillimeter Array (SMA) in Hawai'i. 
Following the correlation, observations were subjected to the standard EHT data reduction path \citep{EHT19b,EHT19c,EHT22b}, including the \texttt{EHT-HOPS} fringe-fitting and post-processing pipeline \citep[][see also \cite{Janssen19} for an alternative pipeline used with the EHT data]{Blackburn19}. 
Additional comments on the data reduction are given in Appendix~\ref{app:AppData}.
The single-dish data used in this paper were observed by the POLAMI \citep{Thum08, Agudo18a} and QUIVER \citep{Myserlis18, Kraus03} programmes on April 4 and April 8, 2017, respectively.

As \C\ exhibits a low jet expansion velocity inside the submilliarcsecond (submas) region, we are able to use quasi-simultaneous VLBI observations of \C\ taken in March and April, 2017, at 15, 43, and 86\,GHz to complement our analysis and assist our interpretation of the underlying jet physics without suffering from time-variability effects.
Here, we define as compact region the entire region probed by the long EHT baselines, with an angular size of smaller than $200\,\upmu\textrm{as}$.
Specifically, we used the publicly available Very Long Baseline Array (VLBA) epochs from April 22, 2017 at 15\,GHz \citep[MOJAVE monitoring program; see][for details regarding the calibration and imaging procedures]{Lister18} and April 16, 2017 at 43\,GHz \citep[VLBA-BU-BLAZAR monitoring program; see][for details regarding the calibration and imaging procedures]{Jorstad17}.
As both monitoring programs publish fully calibrated and imaged data sets, we opted to use them as provided.

\begin{figure}
\centering
\includegraphics[scale=0.45]{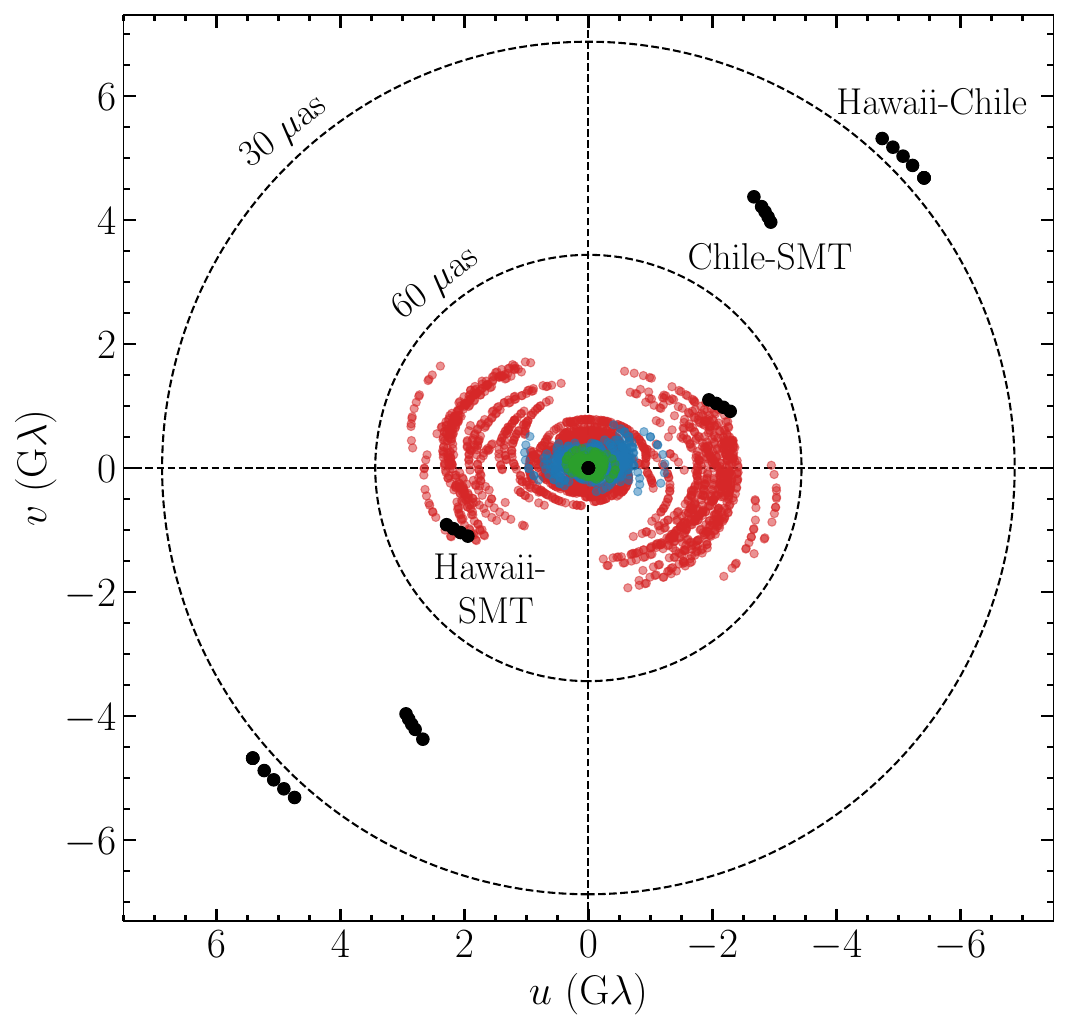}
  \caption{$(u,v)$-coverage of \C, as observed with the VLBA (15\,GHz, green), VLBA (43\,GHz, blue), GMVA (86\,GHz, red), and EHT (228\,GHz, black). 
  Dashed circles indicate fringe spacings characterising the instrumental resolution of 60\,$\upmu\textrm{as}$ and 30\,$\upmu\textrm{as}$. `Chile' denotes stations ALMA and APEX. `Hawaii' denotes stations SMA and JCMT.
  With the higher frequency observations and longer baselines of the EHT, we improve the angular resolution by a factor of $>2$. 
  }
     \label{fig:UV}
\end{figure}

At 86\,GHz, we used the Global Millimeter VLBI Array (GMVA) epoch from March 30, 2017 \citep[see][for details regarding the calibration and imaging procedures]{Paraschos22}.
The antenna instrumental polarisation calibration (D-terms) was performed using the software \texttt{polsolve} \citep{MartiVidal21} and the data were imaged using the \texttt{CLEAN} algorithm \citep[see e.g.][]{Shepherd94, Shepherd95}. The combined $(u, v)$-coverage of our multi-wavelength observations is shown in Fig.~\ref{fig:UV}.

For our analysis we assumed a BH mass of ${M_{\sbullet[1.35]}=8\times10^8\,M_\odot}$ \citep{Scharwaechter13}, for which $100\,\upmu\textrm{as}$ corresponds to ${\sim500\,R_\textrm{s}}$ ($\sim800\,R_\textrm{s}$ deprojected). 
We used $H_0=69.6$ and ${\Omega_\textrm{M} = 0.286}$ in a flat cosmology \citep{Bennet14}, yielding a length scale of $\uppsi=0.36$\,pc/mas. 
Hence, $250\,{\rm R_S}\approx 54\, {\rm \upmu as} \approx 0.02\,{\rm pc}$ ($0.032$\,pc de-projected) at $z=0.0176$.

\subsection{Results}\label{ssec:Results}

We find evidence of a highly ordered, strong magnetic field in the submas compact region of \C.
This region is best fitted by three circular Gaussian components, labelled core (`C'), east (`E'), and west (`W'), as shown in Fig.~\ref{fig:MultiF} (the method we used is described in Appendix~\ref{app:Modelling}). 
The extended flux density detected on the short ALMA-APEX and JCMT-SMA baselines, while resolved out on all long EHT baselines, was fitted by a $\sim5000$ microarcseconds ($\upmu\textrm{as}$) circular Gaussian component with a flux density of $S_\textrm{core}^\textrm{228\,GHz} \sim \Score$.
Furthermore, by averaging\footnote{We used the formula $100\%\times\lvert\sum(Q+\upiota U)\rvert/\sum{I}$, with $I$, $Q$, and $U$ being the image domain Stokes parameters, and performed the pixel-wise summation over the whole field of view.} the linear fractional polarisation measurements of these three components, we determined the net linear fractional polarisation in the compact region to be $m_\textrm{net}=\mtot$.
The short baseline between ALMA and APEX yielded an estimate for the linear fractional polarisation on larger arcsecond scales, of $\sim6\%$ (denoted in the bottom panel of Fig.~\ref{fig:FFF} with the grey marker).

We cross-referenced the submas compact region model fit components at lower frequencies following the method detailed in \cite{Savolainen08} and Appendix~\ref{sec:AppTemplate}. 
While at 15\,GHz the submas region appears fully blended, we are able to recover the 228\,GHz structure at 86\,GHz and even at 43\,GHz. 
The results are reported in Table~\ref{table:Params}. 

We also measured both the total intensity $I$ and linearly polarised emission $P$ in the submas region of \C\ in the 15, 43, and 86\,GHz images. 
The values of linear fractional polarisation at these three lower frequencies are considered as upper limit estimates. 
The results are shown in Fig.~\ref{fig:FFF}.
The VLBI total intensity increases up to the 86\,GHz measurement, and then decreases towards 228\,GHz. 

Close-in-time single-dish measurements at 8, 86, and 228\,GHz are also shown in Fig.~\ref{fig:FFF} (see also Table~\ref{tab:sd}). 
The 86\,GHz flux density is higher than that at 228\,GHz. 
However, the 8\,GHz measurement is also higher than at 86\,GHz, suggesting a significant contribution from the parsec-scale jet. 
Furthermore, at 228\,GHz, the compact-scale VLBI flux density is significantly lower than the corresponding extended flux density, as long EHT baselines over-resolve the large-scale jet emission structure \citep[similar to M\,87, see e.g.][]{EHT19d}. 
In terms of fractional polarisation, it is evident that there is a significant increase at 228\,GHz, indicating a transition in the accretion flow to the optically thin regime.

   \begin{figure*}
   \centering
   \includegraphics[scale=0.14]{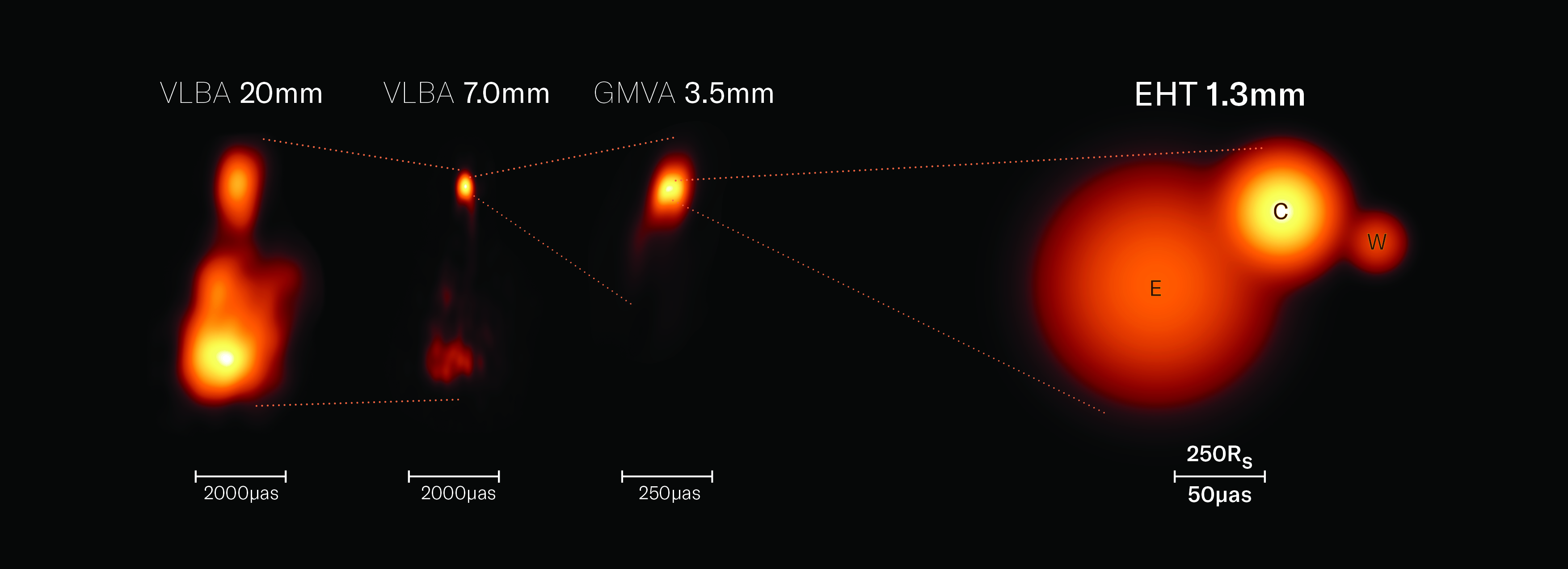}
      \caption{
      Total intensity jet morphology of \C\ at different wavelengths.
      From left to right, we display the 15, 43, 86 (images), and 228\,GHz (model) measurements.
      The horizontal line below each image represents the angular scale. 
      The effective beam sizes, corresponding to these observations are, from left to right, $0.40\times0.60\,\textrm{mas}$, $0.34\times0.16\,\textrm{mas}$, $0.11\times0.04\,\textrm{mas}$, and  $107\times14\,\upmu\textrm{as}$.
      $R_\textrm{S}$ denotes the Schwarzschild radius. 
      }
         \label{fig:MultiF}
   \end{figure*}

   \begin{figure}
   \centering
   \includegraphics[scale=0.45]{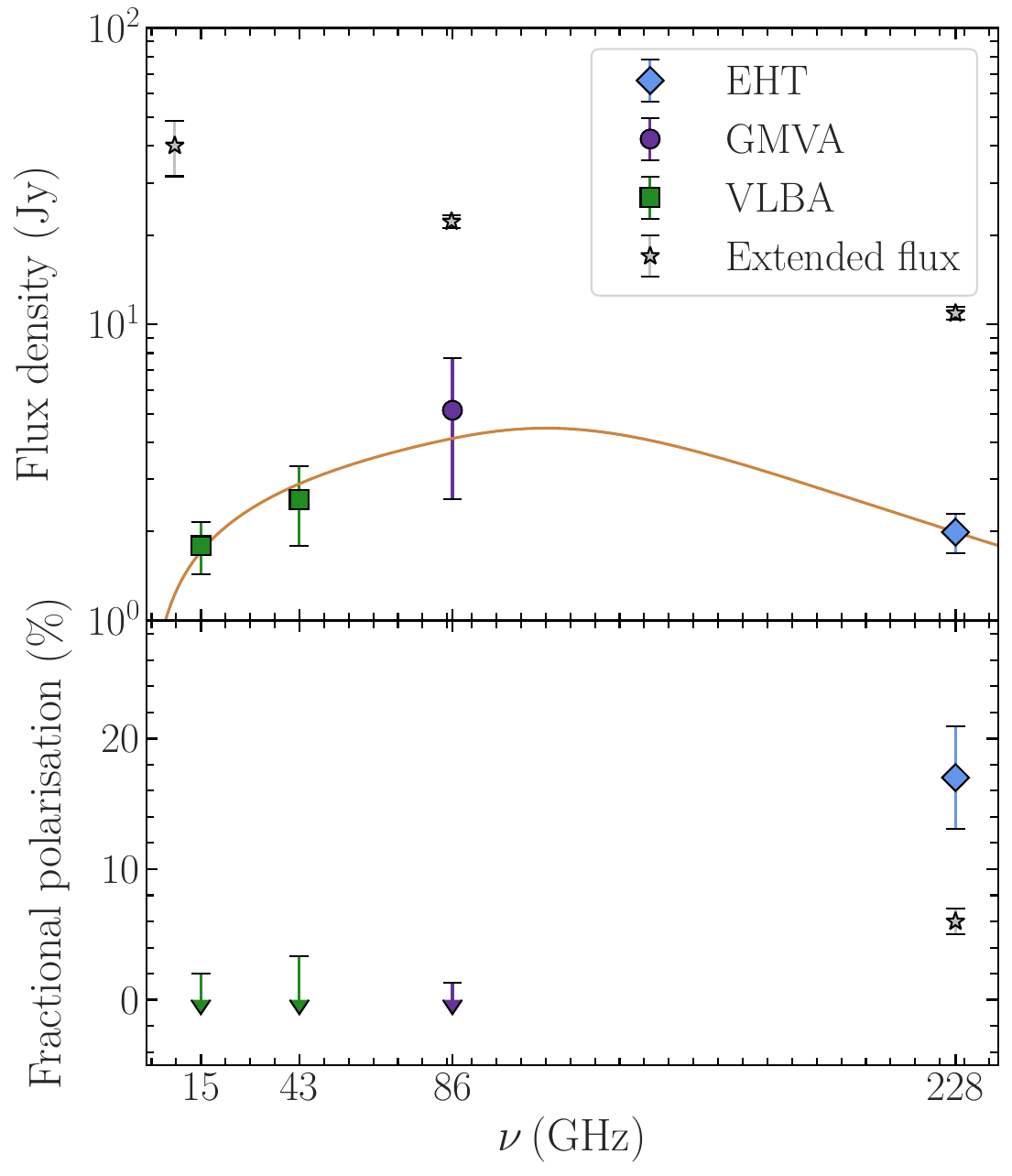}
      \caption{VLBI and single-dish total intensity and fractional polarisation versus frequency for \C, observed in March and April 2017. 
        \emph{Top:} The green box markers, purple dot, and light blue diamond denote the total intensity, compact-scale VLBI flux density measurements at 15, 43, 86, and 228\,GHz, respectively. 
        The latter data point is the sum of components E, C, and W (see Table~\ref{table:Params}).
        The orange line denotes the fit to the spectrum using Eq.~\ref{eq:Turnover}. 
        The grey star markers denote the single-dish (extended) flux density measurements at 8 (QUIVER), 86, and 228\,GHz (POLAMI).
        The turnover frequency is between 86 and 228\,GHz.
        The 8\,GHz single-dish flux density is higher than at 86\,GHz, because the parsec-scale jet flux density contributes to the measurement. 
        \emph{Bottom:} The green and purple arrows indicate the upper limits of the fractional polarisation at 15, 43, and 86\,GHz measured in the same region as the total intensity values in the upper panel.
        The light blue diamond marker again indicates the EHT measurement at 228\,GHz. 
        The grey star marker indicates the zero-baseline fractional polarisation on the baseline between ALMA and APEX. Error bars in both panels indicate the 68\% confidence level.
                }
         \label{fig:FFF}
   \end{figure}

\section{Discussion}\label{sec:Disc}

\subsection{Insights from the synchrotron spectrum}\label{ssec:Core}

Our analysis shows that the east--west elongated core structure \citep{Giovannini18} also persists at 1.3\,mm, and in lower frequencies (as reported at 7\,mm by \citealt{Punsly21} and 3\,mm by \citealt{Oh22}).
Interpretation of the nature of the components comprising this broad core structure heavily depends on the uncertain jet viewing angle ($\upxi$).
An upper limit of $\upxi\sim40^\circ$ was reported by \cite{Oh22} based on a VLBI analysis of the compact region, but much lower values have also been found, for example based on $\gamma$-ray analysis \citep{Abdo09}.
The historically subluminal jet component velocities in the compact region \citep{Punsly21, Hodgson21, Paraschos22} point towards an increased viewing angle.
Moreover, different parts of the jet have been reported to be moving with different velocities, which is related to the so-called `Doppler crisis' phenomenon \citep[e.g.][]{Henri06} and jet stratification \citep{Nagai14}.

The high-resolution, high-frequency EHT observation enables a novel diagnosis of the state of plasma surrounding the central black hole via calculation of the turnover frequency $\upnu_\textrm{m}$ and the synchrotron self-absorption magnetic field strength $B_\textrm{SSA}$. 
Assuming $\upnu_\textrm{m}$ to be 86\,GHz, \cite{Hodgson18} and \cite{Kim19} computed the $B_\textrm{SSA}$ to be $\sim21\,\textrm{G}$.
Using additional EHT flux density measurements, we can directly measure $\upnu_\textrm{m}$. 
While the different observations correspond to different $(u, v)$ coverages, we fitted a focused Gaussian model to the high-signal-to-noise ratio ($S/N$) data at 228\,GHz, finding core diameters within the order of magnitude of the diffraction limit.
We also fixed the sizes of the components for all the frequencies in order to mitigate the effects of the different $(u, v)$ coverages (see Table~\ref{table:Params}).
Subsequently, fitting, then, Eq.~5.90 from \cite{Condon16} (see also \citealt{Rybicki79} and Appendix~\ref{sec:AppMagField}) to the data yields $\upnu_\textrm{m} = \numf$ \citep[see also][]{Tuerler00}.
We computed a core brightness temperature of $T_\textrm{B}=\Tbturn$ from $\upnu_\textrm{m}$, assuming that the angular size of the components at $\upnu_\textrm{m}$ is the same as at 228\,GHz (as the system is optically thin at both frequencies). 
{Within the error budget, the system seems to be in equipartition \citep{Singal86} between kinetic and magnetic energies} \citep[also reported by][based on light-curve variability analysis]{Paraschos23}.

Furthermore, we computed $B_\textrm{SSA} = \Bssa$ using Eq.~2 from \cite{Marscher83} (see also Appendix~\ref{sec:AppMagField}).
We also calculated an equipartition magnetic field strength of $B_\textrm{eq} = \Beq$.
The uncertainties were calculated through standard error propagation.
The two values agree with each other within the error budget.
Our results also tentatively agree within the error budget with the magnetic field reported by \cite{Kim19}. 
The equipartition Doppler factor is $\updelta_\textrm{eq} = \deq$, suggesting that the acceleration happens further downstream, which is in line with lower frequency observations of \C\ \citep[e.g.][and references therein]{Hodgson18, Paraschos22}. 

Moreover, the equipartition magnetic field strength $B_\textrm{eq}$ in the vicinity of the jet apex was computed to reach up to 4\,G in a core shift analysis carried out by \cite{Paraschos21}.
However, the magnetic field value mentioned by these latter authors was calculated at the distance between the extrapolated jet apex and the 86\,GHz core, resulting in a slightly lower estimate than that found in this work.
Nevertheless, it is important to exercise caution when interpreting both $\upnu_\textrm{m}$ and $B_\textrm{SSA}$.
\C\ is a variable source (recently up to 20-30\% variation in total intensity and linear polarised flux density at 43\,GHz within a year based on the monitoring program VLBA-BU-BLAZAR), which means that these observables might be time dependent \citep[compare with the spectrum shown in, e.g.][]{Hodgson18}.
Moreover, our models still contain large uncertainties due to the sparsity of the $(u, v)$ coverage, which may not be fully accounted for.

\subsection{Model interpretation}\label{ssec:Interp}

Possible interpretations of the physical mechanisms driving the wide core structure largely depend on the exact location of the central engine with regard to the observed core.
The current understanding is that the central engine is located north or northwest of the 86\,GHz VLBI core \citep{Giovannini18,Paraschos21}.
As its exact location is still ambiguous, it is unclear whether or not some of the identified components in this work correspond to the core (Case I) or a counter-jet (Case II).

Simulations of the radio jet of M\,87 \citep{Moscibrodzka17} show that the linear polarisation is produced inside the approaching jet, while the dense accretion disc depolarises any radiation reaching us from the counter-jet.
In \C, circumnuclear free-free absorption has already been reported for example by \cite{Walker00}, who cite a possible connection to the accretion disc.
It is thought that the presence of this disc obscures the counter-jet in the milliarcsecond (mas) region of \C, which only becomes visible at a distance of $>2\,\textrm{mas}$ at higher frequencies (as reported e.g. in \citealt{Wajima20} at 86\,GHz).
As both E and W in Fig.~\ref{fig:FIm} are highly linearly polarised (20-80\%), this points towards Case I, meaning that the two components might be at the origin of the double-rail structure seen on larger scales, as opposed to a jet and counter-jet geometry.
However, we note that this interpretation remains speculative, given the uncertainties.

This high fractional polarisation in E and W could be evidence for highly ordered magnetic field lines in the jet plasma with almost no Faraday depolarisation present. 
On the other hand, C has lower fractional polarisation and the synchrotron opacity should be nearly negligible at 228\,GHz according to the Stokes I spectrum shown in Fig~\ref{fig:FFF}.
This may indicate that the main source of depolarisation in the compact region probed by the EHT is beam depolarisation of complex magnetic field patterns or mild Faraday depolarisation, rather than opacity effects.
Consequently, a possible Faraday screen located in the compact region could be at most the size of C, which is $\sim20$\,$\upmu\textrm{as}$.
However, it should be noted here that W is the most uncertain low-total-intensity component, hindering a reliable conclusion about its nature (see also Appendices~\ref{app:AppData} and ~\ref{app:Modelling}).

\C\ is known to show high amounts of Faraday rotation (RM) and the presence of circular polarisation \citep[see e.g. the POLAMI and QUIVER programmes as described in][respectively]{Agudo18a, Myserlis18}.
Using the SMA and CARMA, \cite{Plambeck14} reported an RM of as high as $\sim9\times10^5\,\textrm{rad/m}^2$, indicative of the presence of a strong magnetic field.
This places \C\ in a small group of known radio sources exhibiting similarly high RMs, such as Sgr\,A$^*$ 
($\sim5\times10^5\,\textrm{rad/m}^2$; \citealt{Wielgus22} and references therein), M\,87 ($\sim 10^5\,\textrm{rad/m}^2$; \citealt{Goddi2021}), and PKS\,1830-211 ($\sim10^8\,\textrm{rad/m}^2$; \citealt{MartiVidal15}).
However, whether this RM occurs in the medium surrounding the jet (e.g. from a disc wind) or is connected to the accretion flow remains unknown.
The origin of the RM can be explored by determining its dependence on the observing frequency \citep{Plambeck14, Goddi2021} or the distance from the central engine \citep{Park15}.

The density of the accretion flow, which is a related quantity that can be estimated via the RM, is required in order to constrain the mass-accretion rate around BHs \citep[see][for a relevant discussion about \C]{Nagai17} for different accretion flow models, such as advection-dominated accretion flows (ADAFs; see \citealt{Narayan95}) and convection-dominated accretion flows (CDAFs; see \citealt{Narayan00}).

Different plausible depolarisation mechanisms have been proposed for \C, that is, originating from such an accretion flow and the jet itself \citep{Li16, Kim19}.
Combining the single-dish data presented in Fig.~\ref{fig:FFF}, which were taken quasi-simultaneously with the EHT observations, allows us to estimate an estimate of the RM present in \C. 
We find that $\textrm{RM} = \frm$ by determining the gradient of the EVPAs as a function of the wavelength squared \citep[see also][]{Kim19}. 
The $n\uppi$ ambiguity was resolved beforehand, as described in \cite{Hovatta12}.
Such large RM values could be produced by the presence of relativistic and thermal electrons in the boundary layer between the jet and the interstellar medium, as reported in \cite{Goddi2021} for the jet in M\,87.

\subsection{Physical consequences}\label{ssec:Physics}

The high fractional linear polarisation in the innermost region of \C, revealed at 228\,GHz, clearly indicates that we are probing a previously elusive region, as we are able to achieve higher resolution while being less affected by opacity effects.
We are probing the innermost region of \C\ at $\sim500\,R_\textrm{s}$, which appears to be an optically thin region with an ordered magnetic field framing the core. 

Furthermore, this region is so compact that an association between the broad jet of \C\ and the accretion disk can be ruled out.
However, it should be pointed out that both a BH-driven jet and a disc-driven wind could coexist and the present EHT observations are a better probe of the former.
In a BH-driven jet scenario, jet launching in \C\ might be attributed to a magnetically arrested disc (MAD; see similar simulations carried out for M\,87 in \citealt{Chael19}), as opposed to a thin, broad disc structure \citep{Liska19}. 
Jets in MAD ADAF systems are likely launched by the Blandford-Znajek mechanism \cite{Blandford77}, which is the case where a powerful jet spine is powered directly by the energy extracted from the ergosphere of the BH.

Using our estimate of the RM at 228\,GHz, it is possible to test whether the magnetic field reaches saturation strength; that is, whether the system is in the MAD state \citep{Narayan00, Tchekhovskoy11}. 
Under the assumptions described in Appendix~\ref{sec:AppMagFieldM}, we find that the dimensionless magnetic flux $\upphi = \magpar$ \citep{Tchekhovskoy11}. 
Values above the saturation value $\upphi_\textrm{max} = 50$ indicate that the jet is in a magnetically arrested state, and therefore our analysis suggests that jet launching in \C\ is MAD.
As higher BH spin values and $\upbeta=1.5$ produce values close to $\upphi_\textrm{max}$, our result indicates a preference for a high BH spin and the ADAF model.
\C\ is also classified as a low-luminosity AGN for which ADAF models are commonly invoked \citep{deMenezes20}, further strengthening our conclusion.
The mass-accretion rate estimated in Appendix~\ref{sec:AppMagFieldM} corresponds to $\dot{M} \sim 10^{-5}-10^{-4}\dot{M}_\textrm{Edd}$,
which is somewhat larger than in the case of M\,87 MAD models (see e.g. \citealt{EHT21b}). 
This suggests that a non-negligible dynamical impact of radiation is possible, which could challenge the applicability of the presented analysis.
It should be pointed out here that it is unclear whether Faraday rotation takes place exclusively inside the accretion flow.
Our analysis described in Appendix~\ref{sec:AppMagFieldM} is based on the assumption that the accretion flow is dominant.

If a spine-sheath geometry \citep{Tavecchio14} is present, manifested in the observations as a transverse velocity gradient, it could also be the underlying depolarising structure.
In this case the rotation of the central BH leads to an inhomogeneous and twisted magnetic field topology \cite[see for example][]{Tchekhovskoy15}.
Furthermore, this scenario would also provide an explanation for the Doppler crisis.
As discussed in \cite{Hodgson21}, so-called `jet-in-jet' formations \citep{Giannios09} associated with velocity stratification in the bulk jet flow could be responsible for the enhanced $\gamma$-ray emission observed in \C.  
Such a spine-sheath geometry has already been shown by the EHT to exist on small scales in the jet-launching region of Centaurus\,A \citep{Janssen21}.

Ultimately, our detection of the exceptionally high fractional polarisation at 228\,GHz, the peculiar jet morphology, and the detailed radio spectrum suggest that the jet in \C\ might be launched from both the central BH and the surrounding accretion disc \citep[e.g.][]{Blandford22}.
As shown by the present findings, millimetre VLBI observations pave the way towards probing the ultimate vicinity of BHs. 
Future \C\ EHT observations with added antennas on short and intermediate baselines will help to constrain the jet morphology and improve the fidelity of the model.

\section{Conclusions} \label{sec:Conclusions}

In this work we present the first detection of microarcsecond-scale polarised structures with the EHT.
Our findings can be summarised as follows:
\begin{itemize}
    \item We report the first ever 228\,GHz VLBI model of \C, which reveals that the compact region is made up of three components.
    \item We find a high degree of net fractional polarisation $m_\textrm{net} = \mtot$.
    The brightness temperature is $T_\textrm{b} = \Tbturn$, which suggests that the system is in equipartition.
    \item Using quasi-simultaneous observations of \C\ at 15, 43, 86, and 228\,GHz, we compute a turnover (optical depth $\tau = 1$) frequency of $\nu_\textrm{m} = \numf$, a synchrotron self-absorbed magnetic field of $B_\textrm{SSA} = \Bssa$, and an equipartition magnetic field of $B_\textrm{eq} = \Beq$. 
    However, these values might be influenced by the known variability of the source.
    \item The increased values of linear polarisation suggest that the observed structure is the approaching jet, which is consistent with the large opening angle.
    Such a geometry can be produced by a thick disc associated with a \cite{Blandford77} jet-launching scenario.
    \item We find indications of a preference for higher values of BH spin and the ADAF model in the context of the MAD jet launching prevalent in \C.
\end{itemize}

The EHT is an excellent instrument for probing AGN cores in nearby radio galaxies. 
Combined with lower frequency VLBI arrays, such as the GMVA and the VLBA, the EHT makes it possible to conduct multi-frequency studies, which provide valuable insights into jet formation and jet launching.
New EHT and GMVA observations have already been carried out, with \C\ as the main target.
The increased sensitivity and $(u, v)$ coverage will enable us to conduct follow-up studies with higher fidelity.
Total intensity images of the compact region will shed more light on whether or not the components we were able to identify here correspond to the broad structure seen with {\it RadioAstron} \citep{Giovannini18}.
Spectral index maps of EHT and GMVA images observed quasi-simultaneously might also assist in pinpointing the exact location of the BH \citep[see e.g. Fig.~4 in][]{Paraschos22b} and in discriminating between jet launching-scenarios.

\begin{acknowledgements}
We thank the anonymous referee for the constructive suggestions which improved this manuscript.
The Event Horizon Telescope Collaboration thanks the following
organizations and programs: the Academia Sinica; the Academy
of Finland (projects 274477, 284495, 312496, 315721); the Agencia Nacional de Investigaci\'{o}n 
y Desarrollo (ANID), Chile via NCN$19\_058$ (TITANs) and Fondecyt 1221421, the Alexander
von Humboldt Stiftung; an Alfred P. Sloan Research Fellowship;
Allegro, the European ALMA Regional Centre node in the Netherlands, the NL astronomy
research network NOVA and the astronomy institutes of the University of Amsterdam, Leiden University, and Radboud University;
the ALMA North America Development Fund; the Astrophysics and High Energy Physics programme by MCIN (with funding from European Union NextGenerationEU, PRTR-C17I1); the Black Hole Initiative, which is funded by grants from the John Templeton Foundation and the Gordon and Betty Moore Foundation (although the opinions expressed in this work are those of the author(s) 
and do not necessarily reflect the views of these Foundations); the Brinson Foundation; ``la Caixa'' Foundation (ID 100010434) through fellowship codes LCF/BQ/DI22/11940027 and LCF/BQ/DI22/11940030; 
Chandra DD7-18089X and TM6-17006X; the China Scholarship
Council; the China Postdoctoral Science Foundation fellowships (2020M671266, 2022M712084); Consejo Nacional de Ciencia y Tecnolog\'{\i}a (CONACYT,
Mexico, projects  U0004-246083, U0004-259839, F0003-272050, M0037-279006, F0003-281692,
104497, 275201, 263356);
the Consejer\'{i}a de Econom\'{i}a, Conocimiento, 
Empresas y Universidad 
of the Junta de Andaluc\'{i}a (grant P18-FR-1769), the Consejo Superior de Investigaciones 
Cient\'{i}ficas (grant 2019AEP112);
the Delaney Family via the Delaney Family John A.
Wheeler Chair at Perimeter Institute; Direcci\'{o}n General
de Asuntos del Personal Acad\'{e}mico-Universidad
Nacional Aut\'{o}noma de M\'{e}xico (DGAPA-UNAM,
projects IN112417 and IN112820); 
the Dutch Organization for Scientific Research (NWO) for the VICI award (grant 639.043.513), the grant \newline OCENW.KLEIN.113, and the Dutch Black Hole Consortium (with project No. NWA 1292.19.202) of the research programme the National Science Agenda; the Dutch National Supercomputers, Cartesius and Snellius  
(NWO grant 2021.013); 
the EACOA Fellowship awarded by the East Asia Core
Observatories Association, which consists of the Academia Sinica Institute of Astronomy and
Astrophysics, the National Astronomical Observatory of Japan, Center for Astronomical Mega-Science,
Chinese Academy of Sciences, and the Korea Astronomy and Space Science Institute; 
the European Research Council (ERC) Synergy
Grant ``BlackHoleCam: Imaging the Event Horizon
of Black Holes" (grant 610058); 
the European Union Horizon 2020
research and innovation programme under grant agreements
RadioNet (No. 730562) and 
M2FINDERS (No. 101018682); the Horizon ERC Grants 2021 programme under grant agreement No. 101040021; 
the Generalitat
Valenciana (grants APOSTD/2018/177 and  ASFAE/2022/018) and
GenT Program (project CIDEGENT/2018/021); MICINN Research Project PID2019-108995GB-C22;
the European Research Council for advanced grant `JETSET: Launching, propagation and 
emission of relativistic jets from binary mergers and across mass scales' (grant No. 884631); the FAPESP (Funda\c{c}\~ao de Amparo \'a Pesquisa do Estado de S\~ao Paulo) under grant 2021/01183-8; 
the Institute for Advanced Study; the Istituto Nazionale di Fisica
Nucleare (INFN) sezione di Napoli, iniziative specifiche
TEONGRAV; 
the International Max Planck Research
School for Astronomy and Astrophysics at the
Universities of Bonn and Cologne; 
DFG research grant ``Jet physics on horizon scales and beyond'' (grant No. FR 4069/2-1);
Joint Columbia/Flatiron Postdoctoral Fellowship (research at the Flatiron Institute is supported by the Simons Foundation); 
the Japan Ministry of Education, Culture, Sports, Science and Technology (MEXT; grant JPMXP1020200109); 
the Japan Society for the Promotion of Science (JSPS) Grant-in-Aid for JSPS
Research Fellowship (JP17J08829); the Joint Institute for Computational Fundamental Science, Japan; the Key Research
Program of Frontier Sciences, Chinese Academy of
Sciences (CAS, grants QYZDJ-SSW-SLH057, QYZDJSSW-SYS008, ZDBS-LY-SLH011); 
the Leverhulme Trust Early Career Research
Fellowship; the Max-Planck-Gesellschaft (MPG);
the Max Planck Partner Group of the MPG and the
CAS; the MEXT/JSPS KAKENHI (grants 18KK0090, JP21H01137,
JP18H03721, JP18K13594, 18K03709, JP19K14761, 18H01245, 25120007); the Malaysian Fundamental Research Grant Scheme\newline (FRGS) FRGS/1/2019/STG02/UM/02/6; the MIT International Science
and Technology Initiatives (MISTI) Funds; 
the Ministry of Science and Technology (MOST) of Taiwan (103-2119-M-001-010-MY2, 105-2112-M-001-025-MY3, 105-2119-M-001-042, 106-2112-M-001-011, 106-2119-M-001-013, 106-2119-M-001-027, 106-2923-M-001-005, 107-2119-M-001-017, 107-2119-M-001-020, 107-2119-M-001-041, 107-2119-M-110-005, 107-2923-M-001-009, 108-2112-M-001-048, 108-2112-M-001-051, 108-2923-M-001-002, 109-2112-M-001-025, 109-2124-M-001-005, 109-2923-M-001-001, 110-2112-M-003-007-MY2, 110-2112-M-001-033, 110-2124-M-001-007, and 110-2923-M-001-001);
the Ministry of Education (MoE) of Taiwan Yushan Young Scholar Program;
the Physics Division, National Center for Theoretical Sciences of Taiwan;
the National Aeronautics and
Space Administration (NASA, Fermi Guest Investigator
grant 80NSSC20K1567, NASA Astrophysics Theory Program grant 80NSSC20K0527, NASA NuSTAR award 
80NSSC20K0645); 
NASA Hubble Fellowship 
grants HST-HF2-51431.001-A, HST-HF2-51482.001-A awarded 
by the Space Telescope Science Institute, which is operated by the Association of Universities for 
Research in Astronomy, Inc., for NASA, under contract NAS5-26555; 
the National Institute of Natural Sciences (NINS) of Japan; the National
Key Research and Development Program of China
(grant 2016YFA0400704, 2017YFA0402703, 2016YFA0400702); the National
Science Foundation (NSF, grants AST-0096454,
AST-0352953, AST-0521233, AST-0705062, AST-0905844, AST-0922984, AST-1126433,\newline AST-1140030,
DGE-1144085, AST-1207704, AST-1207730, AST-1207752, MRI-1228509, OPP-1248097, AST-1310896, AST-1440254, 
AST-1555365,\newline AST-1614868, AST-1615796, AST-1715061, AST-1716327,  AST-1716536, OISE-1743747, AST-1816420, AST-1935980, AST-2034306); 
NSF Astronomy and Astrophysics Postdoctoral Fellowship (AST-1903847); 
the Natural Science Foundation of China (grants 11650110427, 10625314, 11721303, 11725312, 11873028, 11933007, 11991052, 11991053, 12192220, 12192223); 
the Natural Sciences and Engineering Research Council of
Canada (NSERC, including a Discovery Grant and
the NSERC Alexander Graham Bell Canada Graduate
Scholarships-Doctoral Program); the National Youth
Thousand Talents Program of China; the National Research
Foundation of Korea (the Global PhD Fellowship
Grant: grants NRF-2015H1A2A1033752, the Korea Research Fellowship Program:\newline
NRF-2015H1D3A1066561, Brain Pool Program: 2019H1D3A1A01102564, 
Basic Research Support Grant 2019R1F1A1059721, 2021R1A6A3A01086420,\newline 2022R1C1C1005255); 
Netherlands Research School for Astronomy (NOVA) Virtual Institute of Accretion (VIA) postdoctoral fellowships; 
Onsala Space Observatory (OSO) national infrastructure, for the provisioning
of its facilities/observational support (OSO receives
funding through the Swedish Research Council under
grant 2017-00648);  the Perimeter Institute for Theoretical
Physics (research at Perimeter Institute is supported
by the Government of Canada through the Department
of Innovation, Science and Economic Development
and by the Province of Ontario through the
Ministry of Research, Innovation and Science); the Princeton Gravity Initiative; the Spanish Ministerio de Ciencia e Innovaci\'{o}n (grants PGC2018-098915-B-C21, AYA2016-80889-P,
PID2019-108995GB-C21, PID2020-117404GB-C21); 
the University of Pretoria for financial aid in the provision of the new 
Cluster Server nodes and SuperMicro (USA) for a SEEDING GRANT approved toward these 
nodes in 2020;
the Shanghai Pilot Program for Basic Research, Chinese Academy of Science, 
Shanghai Branch (JCYJ-SHFY-2021-013);
the State Agency for Research of the Spanish MCIU through
the ``Center of Excellence Severo Ochoa'' award for
the Instituto de Astrof\'{i}sica de Andaluc\'{i}a (SEV-2017-
0709); the Spanish Ministry for Science and Innovation grant CEX2021-001131-S funded by MCIN/AEI/10.13039/501100011033; the Spinoza Prize SPI 78-409; the South African Research Chairs Initiative, through the 
South African Radio Astronomy Observatory (SARAO, grant ID 77948),  which is a facility of the National 
Research Foundation (NRF), an agency of the Department of Science and Innovation (DSI) of South Africa; 
the Toray Science Foundation; the Swedish Research Council (VR); 
the US Department
of Energy (USDOE) through the Los Alamos National
Laboratory (operated by Triad National Security,
LLC, for the National Nuclear Security Administration
of the USDOE, contract 89233218CNA000001); and the YCAA Prize Postdoctoral Fellowship.

We thank
the staff at the participating observatories, correlation
centers, and institutions for their enthusiastic support.
This paper makes use of the following ALMA data:
ADS/JAO.ALMA\#2016.1.01154.V. ALMA is a partnership
of the European Southern Observatory (ESO;
Europe, representing its member states), NSF, and
National Institutes of Natural Sciences of Japan, together
with National Research Council (Canada), Ministry
of Science and Technology (MOST; Taiwan),
Academia Sinica Institute of Astronomy and Astrophysics
(ASIAA; Taiwan), and Korea Astronomy and
Space Science Institute (KASI; Republic of Korea), in
cooperation with the Republic of Chile. The Joint
ALMA Observatory is operated by ESO, Associated
Universities, Inc. (AUI)/NRAO, and the National Astronomical
Observatory of Japan (NAOJ). The NRAO
is a facility of the NSF operated under cooperative agreement
by AUI.
This research used resources of the Oak Ridge Leadership Computing Facility at the Oak Ridge National
Laboratory, which is supported by the Office of Science of the U.S. Department of Energy under contract
No. DE-AC05-00OR22725; the ASTROVIVES FEDER infrastructure, with project code IDIFEDER-2021-086; the computing cluster of Shanghai VLBI correlator supported by the Special Fund 
for Astronomy from the Ministry of Finance in China;  
We also thank the Center for Computational Astrophysics, National Astronomical Observatory of Japan. This work was supported by FAPESP (Fundacao de Amparo a Pesquisa do Estado de Sao Paulo) under grant 2021/01183-8.

APEX is a collaboration between the
Max-Planck-Institut f{\"u}r Radioastronomie (Germany),
ESO, and the Onsala Space Observatory (Sweden). The
SMA is a joint project between the SAO and ASIAA
and is funded by the Smithsonian Institution and the
Academia Sinica. The JCMT is operated by the East
Asian Observatory on behalf of the NAOJ, ASIAA, and
KASI, as well as the Ministry of Finance of China, Chinese
Academy of Sciences, and the National Key Research and Development
Program (No. 2017YFA0402700) of China
and Natural Science Foundation of China grant 11873028.
Additional funding support for the JCMT is provided by the Science
and Technologies Facility Council (UK) and participating
universities in the UK and Canada. 
The LMT is a project operated by the Instituto Nacional
de Astr\'{o}fisica, \'{O}ptica, y Electr\'{o}nica (Mexico) and the
University of Massachusetts at Amherst (USA). The
IRAM 30-m telescope on Pico Veleta, Spain is operated
by IRAM and supported by CNRS (Centre National de
la Recherche Scientifique, France), MPG (Max-Planck-Gesellschaft, Germany), 
and IGN (Instituto Geogr\'{a}fico
Nacional, Spain). The SMT is operated by the Arizona
Radio Observatory, a part of the Steward Observatory
of the University of Arizona, with financial support of
operations from the State of Arizona and financial support
for instrumentation development from the NSF.
Support for SPT participation in the EHT is provided by the National Science Foundation through award OPP-1852617 
to the University of Chicago. Partial support is also 
provided by the Kavli Institute of Cosmological Physics at the University of Chicago. The SPT hydrogen maser was 
provided on loan from the GLT, courtesy of ASIAA.

This work used the
Extreme Science and Engineering Discovery Environment
(XSEDE), supported by NSF grant ACI-1548562,
and CyVerse, supported by NSF grants DBI-0735191,
DBI-1265383, and DBI-1743442. XSEDE Stampede2 resource
at TACC was allocated through TG-AST170024
and TG-AST080026N. XSEDE JetStream resource at
PTI and TACC was allocated through AST170028.
This research is part of the Frontera computing project at the Texas Advanced 
Computing Center through the Frontera Large-Scale Community Partnerships allocation
AST20023. Frontera is made possible by National Science Foundation award OAC-1818253.
This research was done using services provided by the OSG Consortium~\citep{osg07,osg09}, which is supported by the National Science Foundation award Nos. 2030508 and 1836650.
Additional work used ABACUS2.0, which is part of the eScience center at Southern Denmark University. 
Simulations were also performed on the SuperMUC cluster at the LRZ in Garching, 
on the LOEWE cluster in CSC in Frankfurt, on the HazelHen cluster at the HLRS in Stuttgart, 
and on the Pi2.0 and Siyuan Mark-I at Shanghai Jiao Tong University.
The computer resources of the Finnish IT Center for Science (CSC) and the Finnish Computing 
Competence Infrastructure (FCCI) project are acknowledged. This
research was enabled in part by support provided
by Compute Ontario (http://computeontario.ca), Calcul
Quebec (http://www.calculquebec.ca), and Compute
Canada (http://www.computecanada.ca).

The EHTC has received generous donations of FPGA chips from Xilinx
Inc., under the Xilinx University Program. The EHTC
has benefited from technology shared under open-source
license by the Collaboration for Astronomy Signal Processing
and Electronics Research (CASPER). The EHT project is grateful to T4Science and Microsemi for their
assistance with hydrogen masers. This research has
made use of NASA's Astrophysics Data System. We
gratefully acknowledge the support provided by the extended
staff of the ALMA, from the inception of
the ALMA Phasing Project through the observational
campaigns of 2017 and 2018. 
Partly based on observations with the 100-m telescope of the MPIfR (Max-Planck-Institut für Radioastronomie) at Effelsberg.
We would like to thank A. Deller and W. Brisken for EHT-specific support with the use of DiFX. 
We thank Jack Livingston for his comments.
We acknowledge the significance that
Maunakea, where the SMA and JCMT EHT stations
are located, has for the indigenous Hawaiian people.\\

\end{acknowledgements}

\bibliographystyle{aa} 
\bibliography{sources}

\begin{appendix}

\section{Additional methodology comments}\label{app:AppData} 

   \begin{figure*}[h]
   \centering
   \includegraphics[scale=0.4]{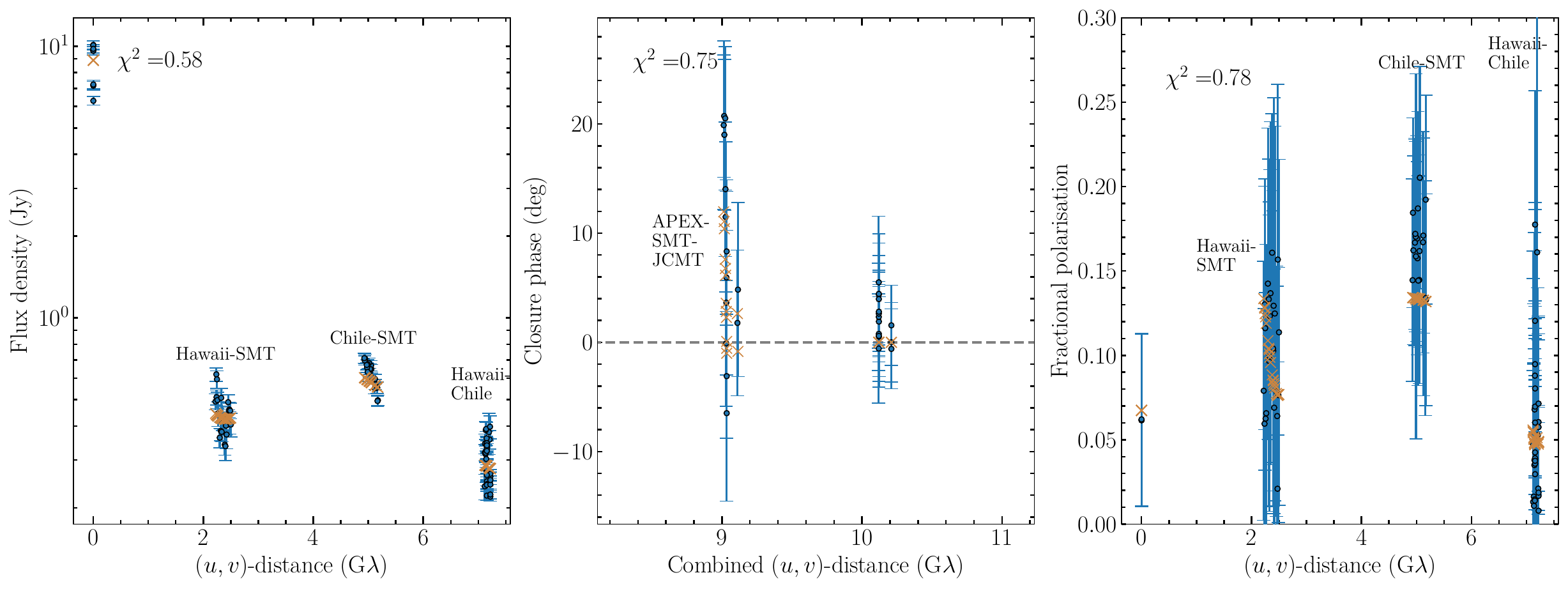}
      \caption{Best-fit model of \C\ compared to the data.
                Presented here from left to right are the data points (denoted with round blue markers) and models (denoted with dark orange crosses) of the visibility amplitudes, closure phases, and fractional polarisation as a function of the $(u, v)$ distance. 
                The combined $(u, v)$ distance used in the middle panel is defined as the square root of the sum of squared lengths of all three baselines forming a triangle. 
                Error bars in all panels indicate the 68\% confidence level.
                }
         \label{fig:Fit}
   \end{figure*}

\begin{table*}
\caption{Summary of the image parameters}
\label{table:Params}
\centering
\begin{tabular}{ccccccc}
Frequency [GHz] & ID & Flux density [Jy] & $T_\textrm{B}$ [$10^{10}\,$K] & FWHM [$\upmu$as] & Polarisation [\%] & Position [$\upmu$as, $\upmu$as]\\
\hline\hline
 15  &  E+C+W &  1.77 &  $253.7 \pm 76.3$  &  $62.1\pm8.1$ &  <2.0           &  (0.0, 0.0)    \\
 \hline
 43  &  E     &  0.07 &  $3.6 \pm 1.5$     &  $35.5\pm7.1$ &  <7.0          &  (34.4, -20.1) \\
     &  C     &  1.69 &  $411.4 \pm 175.8$ &  $16.6\pm3.3$ &  <1.0          &  (0.0, 0.0)    \\
     &  W     &  0.80 &  $535.5 \pm 228.8$ &  $10.0\pm2.0$ &  <2.0          &  (-26.1, -8.58)\\
\hline
 86  &  E     &  1.49 &  $19.8 \pm 8.4$    &  $35.5\pm7.1$ &  <1.0          &  (34.4, -20.1) \\
     &  C     &  3.06 &  $186.2 \pm 79.5$  &  $16.6\pm3.3$ &  <1.0          &  (0.0, 0.0)    \\
     &  W     &  0.58 &  $97.6 \pm 41.7$   &  $10.0\pm2.0$ &  <2.0          &  (-26.1, -8.58)\\
\hline
 228 &  E     &  0.93 &  $1.8 \pm 0.8$     &  $35.5\pm7.1$ &  $20.0\pm5.8$  &  (34.4, -20.1) \\
     &  C     &  1.02 &  $8.8 \pm 3.8$     &  $16.6\pm3.3$ &  $11.0\pm2.0$  &  (0.0, 0.0)    \\
     &  W     &  0.04 &  $1.0 \pm 0.4$     &  $10.0\pm2.0$ &  $40.0-80.0$   &  (-26.1, -8.58)\\
\hline
\end{tabular}\newline

{\raggedright At 15\,GHz the image resolution is insufficient to confidently distinguish between the compact region components, and so we limit ourselves to reporting the integrated flux density and fractional polarisation values instead. 
The positional uncertainty is of the order of $\leq2\%$ for E in the east--west and north--south directions, and $\leq7\%$ in the east--west and $\leq60\%$ in north--south direction for W.
Here, C is fixed at (0, 0). 
The uncertainties of the flux density measurements are of the order of $20\%$ at 15\,GHz, $30\%$ at 43\,GHz, $50\%$ at 86\,GHz, and $15\%$ at 228\,GHz \citep{EHT19c}.  
The large relative uncertainty in the fractional polarisation of W is related to its small total flux density.
The FWHM and positions of the 43, 86, and 228\,GHz components have been fixed in the multi-frequency template-matching framework.
Error margins indicate the 68\% confidence level.\par}
\end{table*}

\begin{table}[! htbp]
\caption{Extended flux measurements}
    \centering
    \begin{tabular}{cc}
    Frequency [GHz] & Flux [Jy]\\
    \hline\hline
         8   & $40.1\,\pm\,8.5$*  \\
         86  & $22.2\,\pm\,1.1$**  \\
         228 & $10.9\,\pm\,0.55$** \\
    \hline
    \end{tabular}\newline

{\raggedright *QUIVER (Effelsberg 100m Telescope)\\ **POLAMI (IRAM 30m Telescope)\par}
    \label{tab:sd}
\end{table}

      \begin{figure}[h]
   \centering
   \includegraphics[scale=0.41]{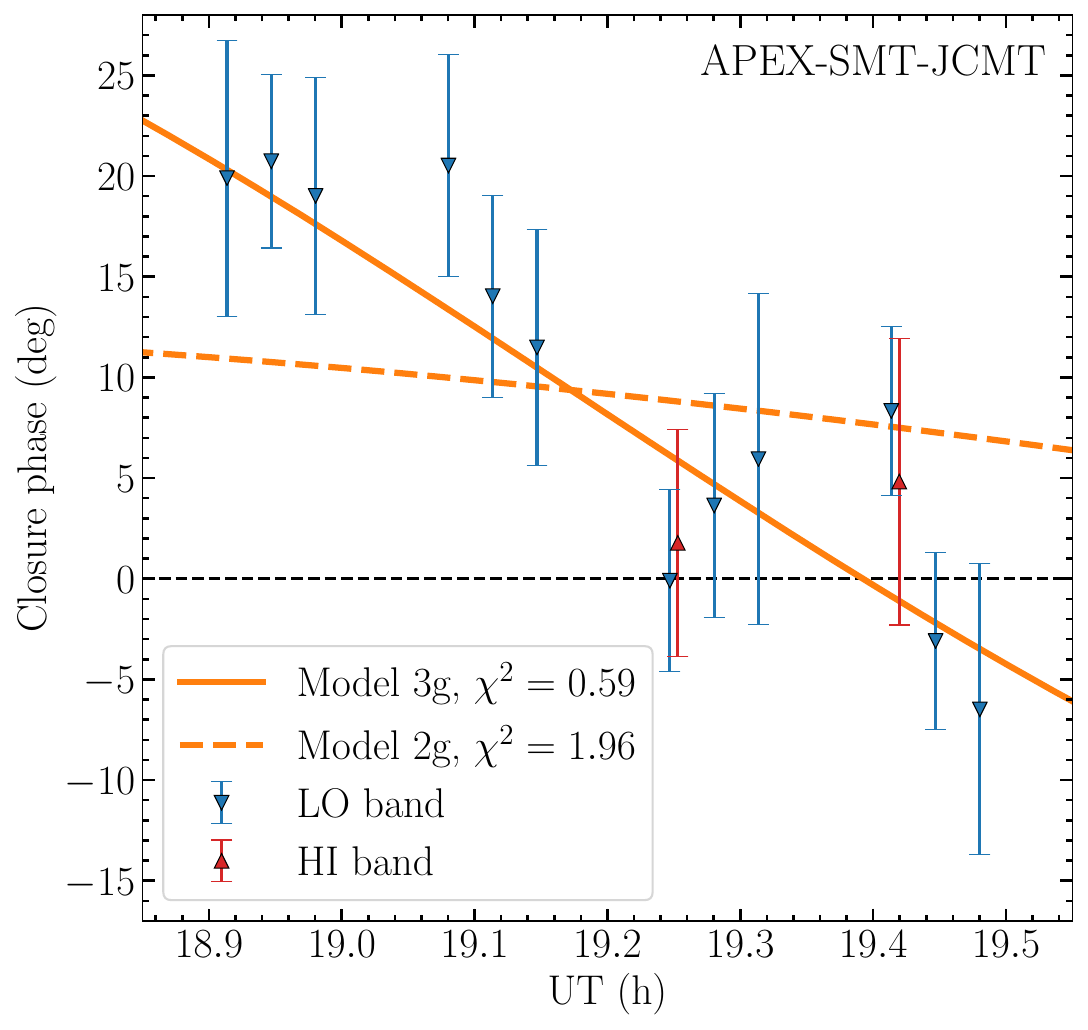}
      \caption{Closure phases as a function of the time of observation, detected on the APEX-SMT-JCMT non-trivial triangle, compared with the predictions of the model presented in this paper. 
      Error bars indicate the 68\% confidence level. 
      The best-fit model 2g with two Gaussian components representing the compact emission region fails to adequately capture the trend in the data, unlike the 3g model with three components. 
      The reported $\chi^2$ corresponds to the non-trivial subset of all measured closure phases. 
                }
         \label{fig:cphases}
   \end{figure}

The frequency setup of the EHT in 2017 consisted of two bands, each of 2\,GHz in width, centred at 227.1\,GHz (LO band) and 229.1\,GHz (HI band): see \cite{EHT19b} for details. 
For all the presented analyses, we combined data sets from both bands, using data points averaged over 2\,GHz in frequency and for 120\,s in time. 
In the EHT data sets, the polarimetric calibration relies on the calibration of ALMA \citep{Goddi19}, which provides the absolute EVPA and serves as a reference site for the computation of the complex polarimetric gains for the entire EHT array \citep{EHT21a}. 
As ALMA was present in only one-sixth of the observing scans of \C, we only obtained the ALMA-based calibration for the single scan. 
The corresponding EVPA measurement on the short ALMA-APEX baseline is close to the north--south axis, consistent with POLAMI observations.
This is in contrast to previous polarimetric analyses with the EHT \citep{EHT21a,Issaoun22,Jorstad2023}, where the absolute EVPA reference was always constrained by ALMA. 
As a consequence, the absolute EVPA calibration was found to be challenging and dependent on additional assumptions for the remaining part of the observations, with the additional difficulties related to APEX drop-outs in the HI band and JCMT observing only the right-hand circular polarisation component, which was used as as proxy for the total intensity in a similar way to in the previous EHT analyses \citep[e.g.][]{EHT19a,EHT22a}.
Hence, in the fitting to the linearly polarised source structure, we made a choice to only fit the absolute values of the fractional Fourier polarisation (corresponding to $|\breve{m}|$, the amplitude ratio of cross-hand and parallel-hand visibility components, following the notation of \citealt{EHT21a}), as shown in the right panel of Fig.~\ref{fig:Fit}. 
We therefore neglect the corrupted linear polarisation phase information.
As a consequence, we only interpret the absolute values of fractional polarisation of the fitted Gaussian components. 
In astrophysical synchrotron plasma, the following order of Stokes parameters magnitude is generally expected: $I>|P| \gg V$; circular polarisation $V$ is consequently neglected in the studies of total intensity I and linear polarisation P. 
Under this assumption, we can use single-polarisation JCMT data in the fitting to absolute values of Fourier fractional linear polarisation.

The high quality of the fit is quantified with the reduced $\chi^2$, which is provided in Fig.~\ref{fig:Fit} for different data products used for the simultaneous fitting: visibility amplitudes, closure phases, and fractional Fourier linear polarisation \citep[see e.g.][for the exact definition]{EHT21a}. 
Additionally, in Fig.~\ref{fig:cphases}, the closure phases observed on the APEX-SMT-JCMT triangle are compared with our model (model 3g, with three Gaussian components representing the compact region emission). 
Non-zero closure phases are an immediate indication of a resolved structure, inconsistent with a circular symmetry of the source \citep{Thompson17}.

The dominant systematic uncertainty in the linear polarisation analysis results comes from the lack of the polarimetric leakage calibration \citep[D-terms calibration; ][]{EHT21a}, which could not be directly employed for the \C\ data set, given the aforementioned data set issues pertaining to the uncertain, time-dependent EVPA calibration. 
In order to obtain a rough characterisation of the impact of the leakage on the resulting polarimetric quantities, we performed a small survey of data sets calibrated with different D-terms. 
We assumed the magnitude of the complex EHT array D-terms estimated and verified in previous EHT publications \citep{EHT21a,Issaoun22,Jorstad2023}, but generated ten random realisations of D-term phases, subsequently refitting the polarimetric source model to data sets with different leakage calibration variants. 
The fractional polarisation uncertainties reported in Table~\ref{table:Params} reflect the results of the D-term calibration survey.

\section{Modelling the EHT data} \label{app:Modelling}

   \begin{figure}
   \centering
   \includegraphics[scale=0.3]{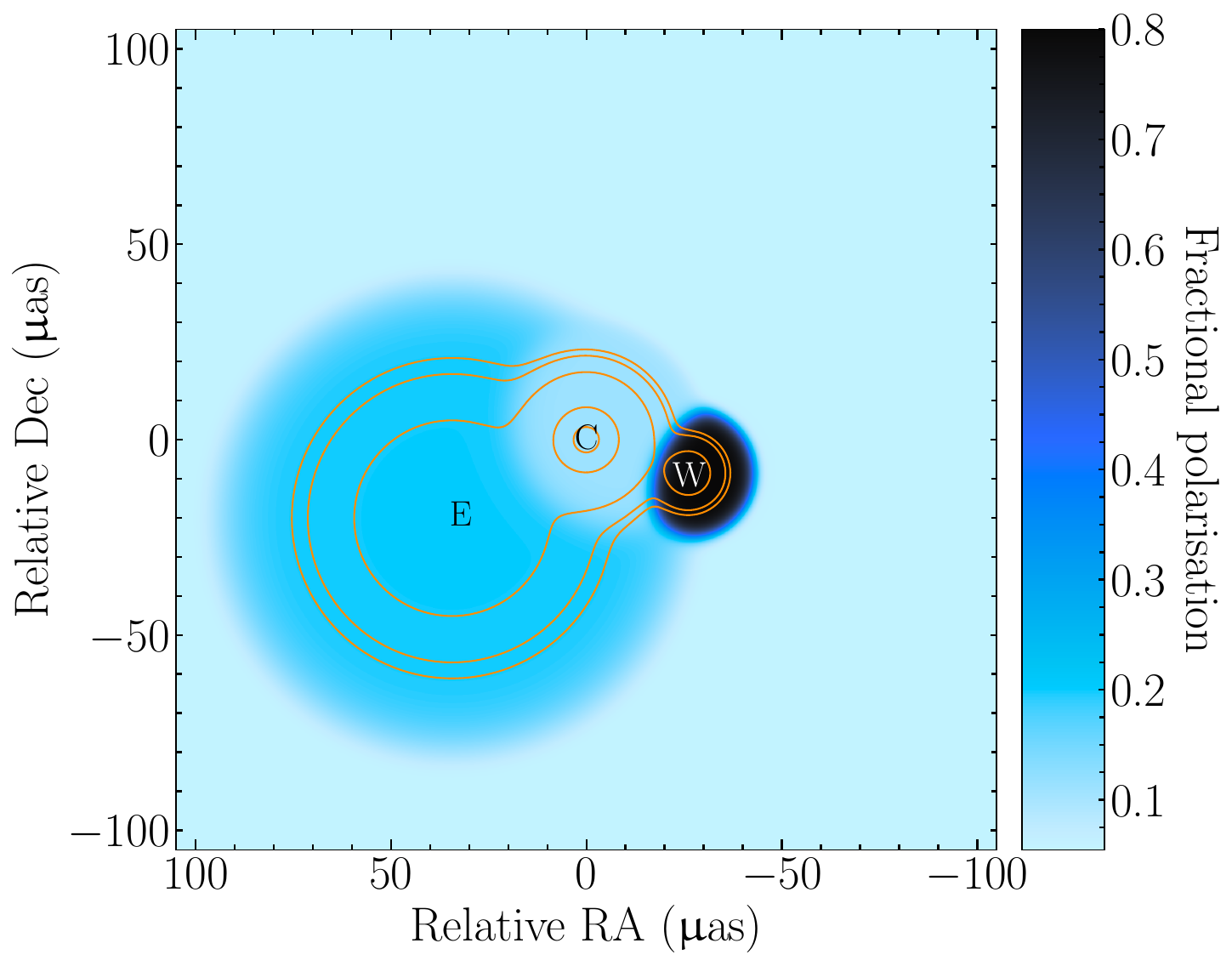}
      \caption{Fractional polarisation image. 
                Shown here is a representation of the best-fit model to the fractional polarisation data in the image plane.
                Contours correspond to 0.5, 1, 5, 50, and 90\% of the peak brightness temperature $T_\textrm{b}=\Tbturnnoe$.
                We note the high net fractional linear polarisation of\, $m_\textrm{net} = \mtot$ in the compact region probed by the EHT.
                }
         \label{fig:FIm}
   \end{figure}

We modelled the EHT data with circular Gaussian components, because the $(u,v)$ coverage is too sparse to reconstruct an image from it. 
Circular instead of elliptical Gaussian components were preferred in order to reduce the number of degrees of freedom.
We used the forward modelling \texttt{eht-imaging} software \citep{Chael16} and leveraged heuristic optimisation tools implemented in \texttt{SciPy} \citep{2020SciPy-NMeth} to search for the best-fit solution in terms of a minimum of an error function.
Figure~\ref{fig:Fit} presents our best-fit model to the visibility amplitudes, closure phases, and fractional polarisation as a function of the $(u,v)$ distance. Figure \ref{fig:FIm} shows the fractional polarisation of the best-fit solution. 
See Appendix~\ref{app:AppData} for additional comments on the fitting procedure.

A certain degree of ambiguity in the compact source structure and the number of components is expected due to the large asymmetry of the EHT $(u,v)$ coverage, which provides resolving power primarily along the south--east/north--west axis; see Fig.~\ref{fig:UV}. 
In particular, we considered the choice between a two-component (2g) and a three-component (3g) model of the compact emission. 
The former corresponds to ten geometric degrees of freedom (i.e., not counting the amplitude gains) for the compact emission and three for the extended emission. 
The latter adds six degrees of freedom to the compact emission part. 
While in the data set used for fitting there are 86 visibility amplitudes, 14 non-trivial closure phases (all shown in Fig.~\ref{fig:cphases}), and 80 absolute values of fractional linear polarisations, there are strong correlations between data points in time, frequency, and location on the $(u,v)$ plane. 
The presence of such correlations reduces the effective constraining power of the data set. 
Hence, the model selection becomes a non-trivial problem. 
The best-fit 2g model essentially recovers the presented geometry of E and C components with the same fractional polarisation of C and around 35\% polarisation of E. 
While the 2g model fits the amplitudes and fractional polarisations well, it is not capable of reproducing the closure phase measurements in a joint fit to all data products, which we demonstrate in Fig.~\ref{fig:cphases}. 
This, along with earlier results obtained at lower observing frequencies \citep{Punsly21}, motivates the selection of the 3g model for the presented analysis, despite it being slightly over-fitted according to the $\chi^2$ values reported in Fig.~\ref{fig:Fit}.
Such behaviour is generally expected for an accurate model given the strong correlations present in the data.

\section{Multi-frequency template matching}\label{sec:AppTemplate} 

We used circular Gaussian components to model the EHT data, determining the number of degrees of freedom required to accurately fit visibility amplitudes, closure phases, and fractional Fourier polarisation, including the modelling of station-based, time-dependent amplitude gains.
This is made possible by using the template-matching technique \citep[e.g.][]{Savolainen08, Kovalev08}, which leverages prior knowledge of the source’s brightness distribution from high-frequency measurements (here 228\,GHz) to estimate the source structure at lower frequencies, even if they are closer together than the lower frequency beam size. 
Given sufficient S/N, structures smaller than the diffraction-limited interferometric resolution can be constrained \cite[e.g.][]{MartiVidal12}. 
For our observations, $S/N\sim550$ at 43\,GHz and $S/N\sim375$ at 86\,GHz, resulting in nominal resolution limits $d_\textrm{lim}^{43\,\textrm{GHz}}=19\,\upmu\textrm{as}$ and $d_\textrm{lim}^{86\,\textrm{GHz}}=6\,\upmu\textrm{as}$, respectively.
Therefore, we were able to apply the high-frequency template, given that the separation between the components comprising the template was sufficiently large.
It should be noted here, that these calculations are performed under the assumption that the source's morphology is a Gaussian. 
However, given the complex structure in the compact region of \C\ revealed here, the actual resolution may be worse.
We therefore adopted the more conservative approach of restricting the resolution limit to the typical value of approximately one-fifth of the beam size \citep[e.g.][]{Oh22}.
This was still possible in our case at 43\,GHz (beam size $\sim100\,\upmu\textrm{as}$) and 86\,GHz (beam size $\sim50\,\upmu\textrm{as}$).
As in our work we investigate the overall spectral behaviour of the submas region, this approach is sufficient to get an estimate for the flux densities and fractional polarisations for each component. 
We also disregarded the core shift effects \citep[see e.g.][]{Paraschos21, Oh22, Paraschos23} between the images at different frequencies, because their effect is negligible for our analysis (of the order of a few tens of $\upmu$as).
Our results are summarised in Table~\ref{table:Params}.
The uncertainties of the flux density measurements are on the order of $20\%$ at 15\,GHz, $30\%$ at 43\,GHz, $50\%$ at 86\,GHz, and $15\%$ at 228\,GHz \citep{EHT19c}.

\section{Magnetic field estimate} 
\label{sec:AppMagField}
We estimated the magnetic field strength in the core via synchrotron turnover frequency fitting. 
The synchrotron spectrum takes the following form \citep{Condon16}: 
\begin{equation}
    S(\upnu) = S_0 \left( \frac{\upnu}{\upnu_\textrm{1}}  \right)^{\upalpha_\textrm{thick}} \left\{  1-\exp \left[ -  \left( \frac{\upnu}{\upnu_\textrm{1}}  \right)^{-(p+4)/2} \right]   \right\}, \label{eq:Turnover}
\end{equation}
for a homogeneous and cylindrical source, where $\upnu_\textrm{1}$ is the frequency where the opacity reaches unity, $\tau=1$, and $S_0=\Szero$ is a multiplication constant, determined from the fit.
Subsequently, $\upnu_\textrm{m}$ is calculated by determining the peak of the fitted spectrum.
Following \cite{Kim19}, we set $\upalpha_\textrm{thick}=0.51\pm0.10$.
The parameter $p$ is the power-law slope of the electron energy distribution function and is set to $p=2$ \citep{Condon16}.

We used two different prescriptions for the magnetic field strength. 
First we calculated the equipartition magnetic field $B_\textrm{eq}$ using \cite{Pacholczyk70} with the following form\footnote{Here, we assume equipartition between cosmic-ray and magnetic energy density.}:
\begin{equation}
    \begin{split}
    \left(\frac{B_\textrm{eq}}{\textrm{G}}\right)&= 2.7\times10^{-7}\\
    &\left[ \frac{\left(1+k_\textrm{u}\right)c_{12}\upkappa_\upnu}{f} \frac{S_\textrm{m}/\textrm{Jy}}{(\uptheta/\textrm{mas})^3}\frac{\upnu_\textrm{m}/\textrm{Hz}}{D_\textrm{L}/\textrm{Gpc}} \frac{\left(1+z\right)^{10}}{\updelta^4} \right]^{2/7}
    \end{split}
     \label{eq:Beq}.
\end{equation}
We note that the exponent $2/7$ only holds for $\upalpha_\textrm{thin} = 0.5$ \citep[see][for a relevant discussion]{Beck05}.
Here, $k_\textrm{u}$ is a ratio that provides an estimate of the energy in relativistic protons compared to electrons and $f$ is a factor denoting the fraction of the total volume of the emitting region occupied by the plasma and magnetic field in equipartition.
Under the assumption of an electron-positron pair plasma \citep[see][for a discussion about electron-positron pair plasma in the vicinity of the SMBH in \C]{Paraschos23}, which is volume filling, $k_\textrm{u} = 0$ and $f = 1$.
The uncertainties for $k_\textrm{u}$ and $f$ are difficult to constrain; their impact on the magnetic field strength computation is discussed below.
The constant $c_{12}$ (in cgs units) is given by the following expression:
\begin{equation}
    c_{12} = c_1^{1/2}c_2^{-1}\left(\frac{2+2\upalpha_\textrm{thick}}{1+2\upalpha_\textrm{thick}}\right)\left(\frac{\upnu_\textrm{min}^{(1+2\upalpha_\textrm{thick})/2}-\upnu_\textrm{max}^{(1+2\upalpha_\textrm{thick})/2}}{\upnu_\textrm{min}^{1+\upalpha_\textrm{thick}} - \upnu_\textrm{max}^{1+\upalpha_\textrm{thick}}}\right),
\end{equation}
where $c_1 = \frac{3e}{4\uppi m^3_\textrm{e}c^5} = 6.27\times10^{18}\,\textrm{[cgs]}$ and $c_2 = \frac{2e^4}{3 m^4_\textrm{e}c^7} = 2.37\times10^{-3}\,\textrm{[cgs]}$; $e$ and $m_\textrm{e}$ are the charge and mass of the electron respectively and $c$ is the speed of light.
Furthermore, $\upkappa_\upnu$ is defined as
\begin{equation}
    \upkappa_\upnu \equiv \frac{\left(\upnu_\textrm{max}/\upnu_\textrm{m}\right)^{1+\upalpha_\textrm{thick}}-1}{1+\upalpha_\textrm{thick}} \frac{\left(\upnu_\textrm{min}/\upnu_\textrm{m}\right)^{1+\upalpha_\textrm{thin}}-1}{1+\upalpha_\textrm{thin}},
\end{equation}
where $\upnu_\textrm{min}$ and $\upnu_\textrm{max}$ are the minimum and maximum frequency range of synchrotron radiation.
We used $\upnu_\textrm{min} = 10^7\,\textrm{Hz}$ (lowest possible frequency for synchrotron emission) and $\upnu_\textrm{max} = 3\times10^{14}\,\textrm{Hz}$ \citep{Biermann87}.
As $[\upnu_\textrm{min}, \upnu_\textrm{max}]$ constitute an assumed frequency range, their values are presented without uncertainties.
The optically thin spectral index is $\upalpha_\textrm{thin}=-0.5$ \citep{Kim19}.
The angular diameter is denoted as $\uptheta$ and the synchrotron peak flux density as $S_{\rm m}=\Sm$ at the frequency $\upnu_{\rm m}$, extrapolated from the optically thin flux density \citep[see also][]{Chamani21}.
We assumed $\updelta=1.1\pm0.1$ for the Doppler factor based on observations presented in \cite{Kim19, Punsly21, Paraschos21} and used $\uptheta=\phim$ (this value already includes the geometric correction discussed in \cite{Marscher83}).

Additionally, using Eq.~2 in \cite{Marscher83}:
\begin{equation}
   \left(\frac{B_\textrm{SSA}}{\rm G}\right) =10^{-50} b(\upalpha_\textrm{thin}) \left(\frac{{\uptheta}}{\rm mas}\right)^4 \left(\frac{\upnu_{\rm m}}{\rm Hz}\right)^5 \left(\frac{S_{\rm m}}{\rm Jy}\right)^{-2} \left(  \frac{\updelta}{1+z} \right).
\end{equation}
Our choice of $\upalpha_\textrm{thin}=-0.5$ results in $b(\upalpha_\textrm{thin})=3.2$.
The resulting estimates for the magnetic field are $B_\textrm{SSA} = \Bssa$ and $B_\textrm{eq} = \Beq$ for the core component C at 228\,GHz.
We point out that the $B_\textrm{SSA}$ calculation is strongly impacted by the value of $\upnu_\textrm{m}$.
An increase or decrease of a few GHz would vary the value of $B_\textrm{SSA}$ by two orders of magnitude.
Similarly, the $B_\textrm{eq}$ calculation strongly depends on the assumption of $k_\textrm{u}$, that is, the particle composition of the jet.
Alternative assumptions of the jet composition resulting in an increase in $k_\textrm{u}$ (diffusive shock acceleration would result in values of $k_\textrm{u}\leq50$, see e.g. \citealt{Bell78}) would increase the value of $B_\textrm{eq}$ by up to a factor of 3.
Likewise, decreasing the value of $f$ (assuming a clumpier medium, filling only half of the total emitting region for example), would result in an increase in $B_\textrm{eq}$ by a factor of $1.2$.
However, we note that the good agreement between the two magnetic field estimates indicates that the choice of $k_\textrm{u} = 0$ and $f = 1$ is reasonable.
The equipartition Doppler factor required for $B_\textrm{SSA}$ to match $B_\textrm{eq}$ is $\updelta_\textrm{eq}=\deq$.

Finally, we can compare $B_\textrm{eq}$ and $B_\textrm{SSA}$ to the strength of the coherent field based on the observed RM.
Using Eq.~15 from \cite{Gardner66}, written:
\begin{equation}
    \textrm{RM} = 8.1\times10^5\int n_\textrm{e} B^\textrm{tot}_\parallel \textrm{d}l,
\end{equation}
we can compute the lower limit of the strength of the ordered field, $B^\textrm{tot}_\parallel$.
Here, $n_\textrm{e}$ is the number density of the thermal electrons, which we set to $n_\textrm{e}=3\times10^4\,\textrm{cm}^{-3}$ \citep{Scharwaechter13}.
The path length of integration through the plasma is $\textrm{d}l$ and can be approximated by $\uppsi\times\uptheta$.
Using these values, $B^\textrm{tot}_\parallel = \Baver$, which is consistent as a lower limit to our calculations of $B_\textrm{eq}$ and $B_\textrm{SSA}$.

\section{Dimensionless magnetic flux} \label{sec:AppMagFieldM}

We calculated the dimensionless magnetic flux $\upphi$ using the expression for the jet power $P_\textrm{jet}$ \cite{Tchekhovskoy10}, which holds for black hole spin values $\upalpha_*\leq1$:
\begin{equation}
    \left(\frac{P_\textrm{jet}}{\dot{M}\textrm{c}^2}\right) = \frac{\upkappa}{4\uppi}\upphi^2\Omega_\textrm{H}^2\left[1+1.38\Omega_\textrm{H}^2-9.2\Omega_\textrm{H}^4\right].
\end{equation}
Here, $\Omega_\textrm{H}\equiv\frac{\lvert\upalpha_*\rvert}{2\left(1+\sqrt{1-\upalpha_*^2}\right)}$, $\dot{M}$ is the mass-accretion rate of \C, and $\upkappa=0.05$ is a constant depending on the initial field geometry.

To determine the mass-accretion rate, we use Eq.~9 from \cite{Marrone06}, in the following form (as also shown in \citealt{Nagai17}):
\begin{equation}
    \begin{split}
    \left(\frac{\dot{M}}{M_\odot\,\textrm{yr}^{-1}}\right)&= 1.3\times10^{-10}\left(1-\left(\frac{r_\textrm{out}}{r_\textrm{in}}\right)^{-(3\upbeta-1)/2}\right)^{-2/3}\\
    &\left(\frac{M_\textrm{BH}}{8.0\times10^8\,M_\odot}\right)^{4/3}
    \left(\frac{2}{3\upbeta-1}\right)^{-2/3}r_\textrm{in}^{7/6}\left(\frac{\textrm{RM}}{\textrm{rad\,m}^{-2}}\right)^{2/3}.\label{eq:mdot}
    \end{split}
\end{equation}
where  $\upbeta=0.5$ for CDAF or $\upbeta=1.5$ for ADAF and $r_\textrm{in}$ and $r_\textrm{out}$ are the inner and outer effective radii of the accretion flow.
Furthermore, $\dot{M}_\textrm{Edd} = L_\textrm{Edd}/(\epsilon c^2)$, where $\epsilon$ is an efficiency factor. 
Using $L_\textrm{Edd} \sim 10^{47}\,\textrm{erg\,s}^{-1}$ \citep{Plambeck14}, $\epsilon=0.1$, $r_\textrm{out}\sim10^5\,R_\textrm{s}$ \citep{Nagai17}, $r_\textrm{in} \sim 90\,R_\textrm{s}$ (size of C), and $\textrm{RM}=\frm$ yields $\dot{M}=0.3\times10^{-3}M_\odot\,\textrm{yr}^{-1}$ for $\upbeta=0.5$ and $\dot{M}\sim1.1\times10^{-3}M_\odot\,\textrm{yr}^{-1}$ $\upbeta=1.5$.
This computation rests on the assumption that the Faraday screen is external.
Our assumption is physically motivated; the jet viewing angle used in our computations ($\upxi\sim40^\circ$) is larger than half of the intrinsic jet opening angle ($i\lesssim20^\circ$; see e.g. \citealt{Paraschos21}), suggesting that we are peering through the jet sheath and boundary layer \citep[see also][]{Plambeck14, Nagai17}.
Furthermore, at 1.3\,mm, we are able to directly examine the environment of the central engine, because the opacity effects become comparatively minor.

Finally, the total luminosity of the jet in \C\ \citep{Rafferty06} is $P_\textrm{jet}=1.5\times 10^{44}\,\textrm{erg\,s}^{-1}$.
Thus, setting $a_*=1$, we computed a range of $\upphi=\magpar$.
Values of $\upphi \gtrsim 50$ refer to MAD models \citep{Tchekhovskoy11, Zamaninasab14}.
We note here that the RM would need to be underestimated by more than an order of magnitude (e.g. due to the Faraday screen not being external) for $\upphi$ to equal its MAD saturation value.
Our investigation of different flow geometries and black hole spins supports an advection-dominated accretion flow in a magnetically arrested state as a model of the jet launching system in the core of \C.

\end{appendix}

\end{document}